\documentclass[12pt]{article}
\usepackage{a4wide,epsfig}
\usepackage{slashed}
\usepackage{amsmath}
\usepackage[small]{caption}
\newcommand{\agt}{\,\rlap{\lower 3.5 pt \hbox{$\mathchar \sim$}} \raise 1pt
 \hbox {$>$}\,}
\newcommand{\alt}{\,\rlap{\lower 3.5 pt \hbox{$\mathchar \sim$}} \raise 1pt
 \hbox {$<$}\,}
\newcommand{\re}{\mathop{\mathrm{Re}}\nolimits}

\catcode`@=11
\newcount\@tempcntc
\def\@citex[#1]#2{\if@filesw\immediate\write\@auxout{\string\citation{#2}}\fi
  \@tempcnta\z@\@tempcntb\m@ne\def\@citea{}\@cite{\@for\@citeb:=#2\do
    {\@ifundefined
       {b@\@citeb}{\@citeo\@tempcntb\m@ne\@citea\def\@citea{,}{\bf ?}\@warning
       {Citation `\@citeb' on page \thepage \space undefined}}%
    {\setbox\z@\hbox{\global\@tempcntc0\csname b@\@citeb\endcsname\relax}%
     \ifnum\@tempcntc=\z@ \@citeo\@tempcntb\m@ne
       \@citea\def\@citea{,}\hbox{\csname b@\@citeb\endcsname}%
     \else
      \advance\@tempcntb\@ne
      \ifnum\@tempcntb=\@tempcntc
      \else\advance\@tempcntb\m@ne\@citeo
      \@tempcnta\@tempcntc\@tempcntb\@tempcntc\fi\fi}}\@citeo}{#1}}
\def\@citeo{\ifnum\@tempcnta>\@tempcntb\else\@citea\def\@citea{,}%
  \ifnum\@tempcnta=\@tempcntb\the\@tempcnta\else
   {\advance\@tempcnta\@ne\ifnum\@tempcnta=\@tempcntb \else \def\@citea{--}\fi
    \advance\@tempcnta\m@ne\the\@tempcnta\@citea\the\@tempcntb}\fi\fi}
\catcode`@=12

\begin{document}

\title{
\vskip-3cm{\baselineskip14pt
\centerline{\normalsize DESY~08--071 \hfill ISSN 0418-9833}
\centerline{\normalsize TTP 08--20\hfill}
\centerline{\normalsize June 2008\hfill}}
\vskip1.5cm
Two-Loop Electroweak Corrections to the $A^0\gamma\gamma$ and $A^0gg$
Couplings of the CP-Odd Higgs Boson}

\author{
{\sc Joachim Brod}\\
{\normalsize Institut f\"ur Theoretische Teilchenphysik, Universit\"at
Karlsruhe}\\
{\normalsize Engesserstra\ss e 7, 76128 Karlsruhe, Germany}\\[1em]
{\sc Frank Fugel}\\
{\normalsize Paul Scherrer Institut,}\\
{\normalsize 5232 Villigen PSI, Switzerland}\\[1em]
{\sc Bernd A. Kniehl}\\
{\normalsize II. Institut f\"ur Theoretische Physik, Universit\"at Hamburg,}\\
{\normalsize Luruper Chaussee 149, 22761 Hamburg, Germany}\\[1em]
}

\date{}

\maketitle


\begin{abstract}
Using the asymptotic-expansion technique, we compute the dominant two-loop
electroweak corrections, of $\mathcal{O}(G_F m_t^2)$, to production and decay
via a pair of photons or gluons of the CP-odd Higgs boson $A^0$ in a
two-Higgs-doublet model with low- to intermediate values of the Higgs-boson
masses and ratio $\tan\beta=v_2/v_1$ of the vacuum expectation values.
We also study the influence of a sequential heavy-fermion generation.
The appearance of three $\gamma_5$ matrices in closed fermion loops requires
special care in the dimensional regularisation of ultraviolet divergences.
The finite renormalisation constant for the pseudoscalar current effectively
restoring the anticommutativity of the $\gamma_5$ matrix, familiar from
perturbative quantum chromodynamics, is found not to receive a correction in
this order.
We also revisit the dominant two-loop electroweak correction to the
$H\to\gamma\gamma$ decay width in the standard model with a fourth fermion
generation.

\medskip

\end{abstract}

\newpage


\section{\label{sec::intro}Introduction}

The search for Higgs bosons is among the prime tasks at the Fermilab Tevatron
and will be so at the CERN Large Hadron Collider (LHC), to go into operation
later during this year, and the International $e^+e^-$ Linear Collider (ILC),
which is currently being designed.
The standard model (SM) contains one complex Higgs doublet, from which
one neutral CP-even Higgs boson ($H$) emerges in the physical particle
spectrum after the electroweak symmetry breaking.
Despite its enormous success in describing almost all experimental particle
physics data available today, the SM is widely believed to be an effective
field theory, valid only at presently accessible energy scales, mainly because
of the naturalness problem related to the fine-tuning of the cut-off scale
appearing quadratically in the Higgs-boson mass counterterm, the failure of
gauge coupling unification, the absence of a concept to incorporate gravity,
and the lack of a cold-dark-matter candidate.
Supersymmetry (SUSY), which postulates the existence of a partner, with spin
shifted by half a unit, to each of the established matter and exchange
particles, is commonly viewed as the most attractive extension of the SM
solving all these problems.
The Higgs sector of the minimal SUSY extension of the SM (MSSM) consists of a
two-Higgs-doublet model (2HDM) and accommodates five physical Higgs bosons:
the neutral CP-even $h^0$ and $H^0$ bosons, the neutral CP-odd $A^0$ boson,
and the charged $H^\pm$-boson pair.
At the tree level, the MSSM Higgs sector has two free parameters, which are
usually taken to be the mass $M_{A^0}$ of the $A^0$ boson and the ratio
$\tan\beta=v_2/v_1$ of the vacuum expectation values of the two Higgs
doublets.

The discovery of the $A^0$ boson would rule out the SM and, at the same time,
give strong support to the MSSM.
At the LHC, this will be feasible except in the wedge of parameter space with
$M_{A^0}\agt250$~GeV and moderate value of $\tan\beta$, where only the $h^0$
boson can be detected \cite{Gianotti:2002xx}.
For low to intermediate values of $\tan\beta$, gluon fusion is by far the
dominant hadroproduction mechanism.
At large values of $\tan\beta$, $A^0b\overline{b}$ associated production
becomes important, too, especially at LHC c.m.\ energy, $\sqrt s=14$~TeV
\cite{Spira:1997dg,Djouadi:2005gj}.
At the ILC operated in the $\gamma\gamma$ mode, via Compton back-scattering of
highly energetic laser light off the electron and positron beams, single
production of the $A^0$ boson will allow for its discovery, also throughout a
large fraction of the LHC wedge, and for a precision determination of its
profile \cite{Muhlleitner:2001kw}.
Two-photon collisions, albeit with less luminosity, will also take place in
the regular $e^+e^-$ mode of the ILC through electromagnetic bremsstrahlung or
beamstrahlung off the lepton beams.

In the mass range $M_{A^0}<2m_t$ and for large values of $\tan\beta$ in the
whole $M_{A^0}$ range, the $A^0$ boson dominantly decays to a $b\overline{b}$
pair, with a branching fraction of about 90\%
\cite{Spira:1997dg,Djouadi:2005gj}.
As in the case of the $H$ boson of the SM, the rare $\gamma\gamma$ decay
channel then provides a useful signature at the LHC if the $b$ and
$\overline{b}$ quarks cannot be separated sufficiently well from the
overwhelming background from quantum chromodynamics (QCD).
The $A^0\to gg$ channel will greatly contribute to the decay mode to a
light-hadron dijet, which will be measurable at the ILC.

Since the $A^0$ boson is neutral and colourless, the $A^0\gamma\gamma$ and
$A^0gg$ couplings are loop induced.
As the $A^0$ boson has no tree-level coupling to the $W$ boson and its
coupling to sfermions flips their ``handedness'' (left or right), the
$A^0\gamma\gamma$ coupling is mediated at leading order (LO) by heavy quarks
and charged leptons and by light charginos, whereas heavy charginos decouple
\cite{Kalyniak:1985ct}.
The $A^0gg$ coupling is generated at LO by heavy-quark loops
\cite{Gunion:1986nh}.

Reliable theoretical predictions for the $A^0\gamma\gamma$ and $A^0gg$
couplings, including higher-order radiative corrections, are urgently required
to match the high precision to be reached by the LHC and ILC experiments
\cite{Battaglia:2000jb,Weiglein:2004hn}.
Specifically, the properties of the $A^0$ boson, especially its CP-odd nature,
must be established, and the sensitivity to novel high-mass particles
circulating in the loops must be optimised.
The present state of the art is as follows.
The next-to-leading-order (NLO) QCD corrections,
of relative order ${\cal O}(\alpha_s)$ in the strong-coupling constant
$\alpha_s$, to the partial decay widths $\Gamma(A^0\to\gamma\gamma)$
\cite{Djouadi:1993ji,Spira:1995rr} and $\Gamma(A^0\to gg)$
\cite{Spira:1995rr}, and the production cross section $\sigma(gg\to A^0)$
\cite{Spira:1995rr,Spira:1993bb} are available for arbitrary values of quark
and $A^0$-boson masses as one-dimensional integrals, which were solved in
terms of harmonic polylogarithms for $\Gamma(A^0\to\gamma\gamma)$,
$\Gamma(A^0\to gg)$, and the virtual correction to $\sigma(gg\to A^0)$
\cite{Harlander:2005rq,Aglietti:2006tp}.
The latter was also obtained for general colour factors of the gauge group
SU($N_c$) in the limit $m_t\to\infty$ using an effective Lagrangian
\cite{Ravindran:2004mb}.
In the same way, the ${\cal O}(\alpha_s)$ correction to $\sigma(gg\to A^0)$
was first calculated in Ref.~\cite{Kauffman:1993nv}.

The next-to-next-to-leading-order (NNLO) QCD corrections, of
${\cal O}(\alpha_s^2)$, to $\Gamma(A^0\to gg)$ \cite{Chetyrkin:1998mw}
and $\sigma(gg\to A^0)$ \cite{Harlander:2002vv} were found for $m_t\to\infty$
using an effective Lagrangian.
The ${\cal O}(\alpha_s)$ SUSY QCD correction, due to virtual squarks and
gluinos besides the heavy quarks, to $\sigma(gg\to A^0)$ was obtained from
an effective Lagrangian constructed by also integrating out the SUSY particles
\cite{Harlander:2005if}.
The two-loop master integrals appearing in the latter calculation if the
masses of the virtual scalar bosons and fermions are kept finite were
expressed in terms of harmonic polylogarithms \cite{Aglietti:2006tp}.

In this paper, we take the next step and present the dominant electroweak
corrections to $\Gamma(A^0\to\gamma\gamma)$ and $\Gamma(A^0\to gg)$ at NLO.
Our key results were already summarised in Ref.~\cite{Brod:2008ct}.
Here, we present the full details of our calculation and a comprehensive
discussion of its phenomenological implications.
Since these corrections are purely virtual, arising from two-loop diagrams,
they carry over to $\sigma(\gamma\gamma\to A^0)$ and $\sigma(gg\to A^0)$, via
\begin{equation}
\sigma(\gamma\gamma/gg\to A^0)=\frac{8\pi^2}{N_{\gamma,g}^2M_{A^0}}
\Gamma(A^0\to\gamma\gamma/gg)\delta\left(\hat{s}-M_{A^0}^2\right),
\end{equation}
where $N_\gamma=1$ and $N_g=N_c^2-1=8$ are the colour multiplicities of the
photon and the gluon, respectively, and $\hat{s}$ is the partonic c.m.\ energy
square.
For the time being, we focus our attention on the particularly interesting
region of parameter space with low to intermediate Higgs-boson masses,
$M_{h^0},M_{H^0},M_{A^0},M_{H^\pm}<m_t$,\footnote{%
As for $M_{A^0}$, we actually need $M_{A^0}<2M_{W^\pm},2M_{H^\pm}$ in order
for asymptotic expansion to be applicable.}
and low to moderate value of
$\tan\beta$, $\tan\beta\ll m_t/m_b$, and assume that the SUSY particles are so
heavy that they can be regarded as decoupled, yielding subdominant
contributions.
The dominant electroweak two-loop corrections are then induced by the top
quark and are of relative order ${\mathcal O}(x_t)$, where
$x_t=G_Fm_t^2/(8\pi^2\sqrt{2})\approx3.17\times10^{-3}$ with $G_F$ being
Fermi's constant.
We also consider the influence of a sequential generation of heavy fermions
$F$, beyond the established three generations, which generate corrections of
generic order ${\mathcal O}(x_F)$.
Such corrections were already studied for the $H\to\gamma\gamma$ decay in the
SM supplemented with a fourth fermion generation in
Ref.~\cite{Djouadi:1997rj}, and we revisit this analysis.

In the calculation of two-loop electroweak corrections to the
$A^0\gamma\gamma$ and $A^0gg$ couplings, one encounters closed fermion loops
involving three $\gamma_5$ matrices, so that the use of the na\"\i ve
anticommuting definition of the $\gamma_5$ matrix is bound to fail.
This leads us to employ the 't~Hooft-Veltman-Breitenlohner-Maison (HVBM)
\cite{tHooft:1972fi} scheme and a finite renormalisation constant, $Z_5^p$,
for the pseudoscalar current to effectively restore the anticommutativity of
the $\gamma_5$ matrix \cite{Trueman:1979en,Collins:1984xc,Larin:1993tq}, which
is so far only known within QCD \cite{Larin:1993tq}.

This paper is organised as follows.
In Section~\ref{sec::method}, we explain our method of calculation and
evaluate $Z_5^p$ to ${\mathcal O}(x_t)$ and ${\mathcal O}(x_F)$.
In Section~\ref{sec::pseudoscalar}, we calculate the ${\mathcal O}(x_t)$ and
${\mathcal O}(x_F)$ corrections to $\Gamma(A^0\to\gamma\gamma)$ and
$\Gamma(A^0\to gg)$ in the 2HDM with three and four fermion generations.
We also recover the ${\mathcal O}(\alpha_s)$ correction to
$\Gamma(A^0\to\gamma\gamma)$.
In Section~\ref{sec::scalar}, we recalculate the ${\mathcal O}(x_F)$ correction
to $\Gamma(H\to\gamma\gamma)$ due to a fourth fermion generation added on top
of the SM.
In Section~\ref{sec::numerics}, we present our numerical results.
We conclude with a summary in Section~\ref{sec::summary}.


\section{\label{sec::method}Method of calculation}

As explained below, we assume a strong hierarchy between the heavy fermions on
the one hand and the gauge and Higgs bosons on the other hand, so that we may
extract the leading corrections using the asymptotic-expansion technique
\cite{Smirnov:pj}.
We use a completely automated set-up, which relies on the successive use of
the computer programs \texttt{QGRAF} \cite{Nogueira:1991ex}, \texttt{q2e},
\texttt{exp} \cite{Harlander:1997zb}, and \texttt{MATAD}
\cite{Steinhauser:2000ry}.
First, \texttt{QGRAF} is used to generate the Feynman diagrams.
Its output is then rewritten by \texttt{q2e} to be understandable by
\texttt{exp}.
The latter performs the asymptotic expansion and generates the relevant
subdiagrams according to the rules of the so-called hard-mass procedure
\cite{Smirnov:pj}.
\texttt{Form} \cite{Vermaseren} files are generated.
They can be read by \texttt{MATAD}, which performs the calculation of the
diagrams.

Since we consider the SUSY partners to be decoupled, we may as well work in
the 2HDM without SUSY.
We may thus extract the ultraviolet (UV) divergences by means of dimensional
regularisation, with $d=4-2\epsilon$ space-time dimensions and 't~Hooft mass
scale $\mu$, without introducing SUSY-restoring counterterms
\cite{Hollik:2001cz}.
For convenience, we work in 't~Hooft-Feynman gauge.
We take the Cabibbo-Kobayashi-Maskawa quark mixing matrix to be unity, which
is well justified because the third quark generation is, to good
approximation, decoupled from the first two \cite{Yao:2006px}.
We adopt Sirlin's formulation of the electroweak on-shell renormalisation
scheme \cite{Sirlin:1980nh}, which uses $G_F$ and the physical particle masses
as basic parameters, and its extension to the MSSM \cite{Hollik:2002mv}.
Various prescriptions for the renormalisation of the auxiliary variable
$\tan\beta$, with specific virtues and flaws, may be found in the literature,
none of which is satisfactory in all respects (for a review, see
Ref.~\cite{Freitas:2002um}).
For definiteness, we employ the Dabelstein-Chankowski-Pokorski-Rosiek (DCPR)
scheme \cite{Chankowski:1992er,Dabelstein:1995js}, which maintains the
relation $\tan\beta=v_2/v_1$ in terms of the ``true'' vacua through the
condition $\delta v_1/v_1=\delta v_2/v_2$, and demands the residue condition
$\re\hat\Sigma_{A^0}^\prime(M_{A^0})=0$ and the vanishing of the $A^0$--$Z^0$
mixing on shell as $\re\hat\Sigma_{A^0Z^0}(M_{A^0})=0$, where
$\hat\Sigma_{A^0}(q^2)$ and $\hat\Sigma_{A^0Z^0}(q^2)$ are the renormalised
$A^0$-boson self-energy and $A^0$--$Z^0$ mixing amplitude, respectively.
It has been pointed out \cite{Freitas:2002um} that the DCPR definition of
$\tan\beta$ is gauge dependent.
We do not actually encounter this drawback in our analysis, since we need to
renormalise $\tan\beta$ to ${\mathcal O}(x_f)$, so that only fermion loops
contribute.
However, this problem will show up when subleading terms of the two-loop
electroweak corrections are to be computed.
Our final results can be straightforwardly converted to any other
renormalisation prescription for $\tan\beta$, by substituting the finite
relationship between the old and new definitions of $\tan\beta$.

As already mentioned in Section~\ref{sec::intro}, the evaluation of the
relevant two-loop diagrams is aggravated by the appearance of three $\gamma_5$
matrices inside closed fermion loops.
This leads us to adopt the HVBM scheme \cite{tHooft:1972fi}, which allows for
a consistent treatment of the Dirac algebra within the framework of
dimensional regularisation.
In this scheme, the $\gamma_5$ matrix is given by
\begin{equation}
\gamma_5=\frac{i}{4!}\varepsilon_{\mu\nu\rho\sigma}\,
\gamma^\mu\gamma^\nu\gamma^\rho\gamma^\sigma,
\end{equation}
where the totally antisymmetric Levi-Civita tensor is defined in $d$
dimensions as
\begin{equation}
\varepsilon_{\mu\nu\rho\sigma}=\begin{cases}
1 & \text{if $(\mu,\nu,\rho,\sigma)$ even permutations of (0,1,2,3),}\\
-1 & \text{if $(\mu,\nu,\rho,\sigma)$ odd permutations of (0,1,2,3),}\\
0 & \text{otherwise.}\\
\end{cases}
\label{eq:epsilon}
\end{equation}
This definition leads to the following mixed anticommutation and commutation
relations, where we have to distinguish between 4 and $(d-4)$ dimensions:
\begin{eqnarray}
\{\gamma^\mu,\gamma_5\}&=&0,\quad\mbox{if $\mu=0,1,2,3$,}
\nonumber\\
\left[\gamma^\mu,\gamma_5\right]&=&0,\quad\mbox{otherwise.}
\end{eqnarray}
By giving up the full anticommutation property of $\gamma_5$, we can retain
the familiar expression for the trace of four $\gamma$ matrices and one
$\gamma_5$ matrix:
\begin{equation}
\text{tr}(\gamma_\mu\gamma_\nu\gamma_\rho\gamma_\sigma\gamma_5)
=4i\varepsilon_{\mu\nu\rho\sigma}.
\end{equation}
Traces involving less than four $\gamma$ matrices and one $\gamma_5$ matrix
vanish.
We introduce the following projectors onto the 4- and $(d-4)$-dimensional
subspaces:
\begin{eqnarray}
\tilde{g}^{\mu\nu}&=&\begin{cases}
g^{\mu\nu} & \mbox{if $\mu$ and $\nu$ are smaller than 4,}\\
0 & \text{otherwise;}\\
\end{cases}\nonumber\\
\hat{g}^{\mu\nu}&=&\begin{cases}
g^{\mu\nu} & \mbox{if $\mu$ and $\nu$ are larger than 3,}\\
0 & \text{otherwise.}\\
\end{cases}
\end{eqnarray}
Here and in the following, we label quantities in 4 dimensions with a tilde,
quantities in ($d-4$) dimensions with a hat, and quantities in $d$ dimensions
without superscript.
For a given four-vector $V$, we thus have
\begin{eqnarray}
\tilde{V}^{\mu}&=&\tilde{g}^{\mu\nu}V_{\nu},
\nonumber\\
\hat{V}^{\mu}&=&\hat{g}^{\mu\nu}V_{\nu}.
\end{eqnarray}
The projectors fulfil the following relations:
\begin{eqnarray}
\tilde{g}^{\mu}_{\mu}=g^{\mu\nu}\tilde{g}_{\mu\nu}
&=&\tilde{g}^{\mu\nu}\tilde{g}_{\mu\nu}=4,
\nonumber\\
\hat{g}^{\mu}_{\mu}=g^{\mu\nu}\hat{g}_{\mu\nu}
&=&\hat{g}^{\mu\nu}\hat{g}_{\mu\nu}=d-4,
\nonumber\\
\tilde{g}^{\mu\nu}\hat{g}_{\mu\nu}
&=&0.
\end{eqnarray}
More details may be found in Ref.~\cite{Collins:1984xc}.
We explicitly verified that, in our case, the na\"\i ve anticommuting
definition of the $\gamma_5$ matrix yields ambiguous results, which depend on
the way of executing the Dirac traces.

The actual implementation of these rules into the \texttt{MATAD} setup is
accomplished in two independent ways.
Firstly, we use the \texttt{Mathematica} package \texttt{TRACER}
\cite{Jamin:1991dp} and compute the traces separately.
Secondly, we implement our own \texttt{FORM} routine for evaluating the
traces in the HVBM scheme.
Both methods yield the same results. 

The application of the HVBM scheme introduces loop momenta that are projected
onto the ($d-4$)-dimensional subspace.
These have to be expressed through loop momenta in the full $d$-dimensional
space because \texttt{MATAD} performs the integration in $d$ dimensions.
For instance, in the case of a massive one-loop tadpole with loop momentum
$q$, we substitute
\begin{equation}
\hat{q}^2=\frac{4}{d}q^2
\end{equation}
in the numerator of the integrand.
Similar identities can be derived for all the other cases.

\begin{figure}[ht]
\begin{center}
\includegraphics[width=0.75\textwidth,viewport=127 695 471 752,clip]{%
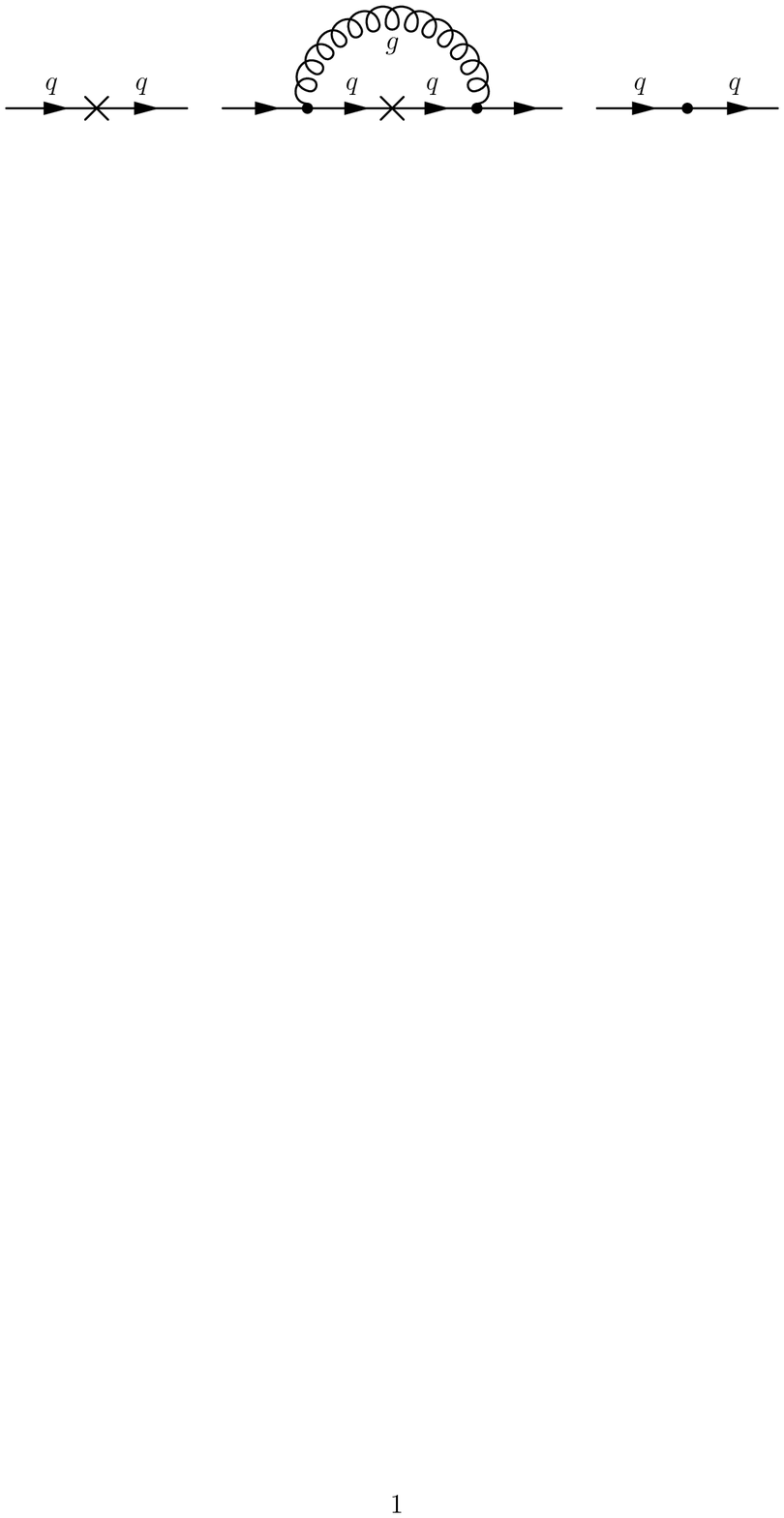}
\centerline{\hspace{3.2cm}(a)\hfill(b)\hfill(c)\hspace{3.2cm}}
\caption{\label{fig:pscala}%
Feynman diagrams contributing to $Z_5^p$ at ${\cal O}(\alpha_s)$.
Crosses and dots indicate insertions of $P(x)$ and its operator
renormalisation $Z_2Z^pZ_5^p$, respectively.}
\end{center}
\end{figure}

Furthermore, we have to introduce an additional finite counterterm, $Z_5^p$,
in the renormalisation of the pseudoscalar current to effectively restore the
anticommutativity of the $\gamma_5$ matrix
\cite{Trueman:1979en,Collins:1984xc,Larin:1993tq}.
Within QCD, $Z_5^p$ is known through ${\cal O}(\alpha_s^3)$
\cite{Larin:1993tq}.
Here, we need $Z_5^p$ at ${\mathcal O}(x_t)$ and ${\mathcal O}(x_F)$.
In order to explain our procedure, we first repeat the derivation of the
${\cal O}(\alpha_s)$ term.
For simplicity, we work in massless QCD with 't~Hooft-Feynman gauge.
As usual, we adopt the modified minimal-subtraction ($\overline{\mathrm{MS}}$)
renormalisation scheme.  
The pseudoscalar current is defined in coordinate space as 
\begin{equation}
P(x)=Z_2Z^pZ_5^p\bar{\psi}(x)\gamma_5\psi(x),
\end{equation}
where $Z_2$ and $Z^p$ are the usual wave-function and pseudoscalar-current
renormalisation constants, respectively.
Passing to momentum space, we define
\begin{equation}
\mathrm{i}S_F(p)\Gamma_5(p,p^\prime)\mathrm{i}S_F(p^\prime)
=\int\mathrm{d}^dx\mathrm{d}^dy\,
\mathrm{e}^{\mathrm{i}(p\cdot x-p^\prime\cdot y)}
\langle0|T[\psi(x)P(0)\overline{\psi}(y)]|0\rangle,
\end{equation}
where $T$ denotes the time-ordered product and
$iS_F(p)=\int\mathrm{d}^dx\,\mathrm{e}^{\mathrm{i}p\cdot x}\langle0|T[\psi(x)
\overline{\psi}(0)]|0\rangle$
is the Feynman propagator of the quark.
The key quantity for our purposes is then the amputated Green function
$\Gamma_5(p,p)$ at zero momentum transfer.
The Feynman diagrams relevant through ${\cal O}(\alpha_s)$ are depicted in
Fig.~\ref{fig:pscala}, where crosses and dots indicate the insertions of
$P(x)$ and the operator renormalisation, respectively, and it is
understood that external legs are amputated.
The tree-level diagram in Fig.~\ref{fig:pscala}(a) yields
\begin{equation}
\Gamma_5^{(0)}(p,p)=\gamma_5.
\label{eq:ptree}
\end{equation}
The one-loop diagram in Fig.~\ref{fig:pscala}(b) leads to the integral
\begin{equation}
\Gamma_5^{(1)}(p,p)
= 4 \pi \alpha_s C_F \int\frac{\mathrm{d}^d q}{(2\pi)^d}\,
\frac{-i\gamma_{\mu}\slashed{q}\gamma_5\slashed{q}\gamma^{\mu}}
{(p-q)^2(q^2)^2}, 
\end{equation}
where $C_F=(N_c^2-1)/(2N_c)=4/3$ is the eigenvalue of the Casimir operator of
the fundamental representation of SU($N_c$), with $N_c=3$ for QCD.
We decompose the string of gamma matrices in the numerator as 
\begin{equation}
\gamma_{\mu}\slashed{q}\gamma_5\slashed{q}\gamma^{\mu}
=\gamma_{\mu}\slashed{q}\slashed{q}\gamma^{\mu}\gamma_5 
+\gamma_{\mu}\slashed{q}(-2\slashed{q}\hat{\gamma}^{\mu}
-2\slashed{\hat{q}}\gamma^{\mu} 
+4\slashed{\hat{q}}\hat{\gamma}^{\mu})\gamma_5.
\label{eq:pnum}
\end{equation}
The first term on the right-hand side of Eq.~(\ref{eq:pnum}) is what we would
obtain using an anticommuting $\gamma_5$ matrix.
Upon loop integration it yields an expression proportional to
Eq.~(\ref{eq:ptree}), the divergent part of which is 
\begin{equation}
\Gamma_5^{(1),\,\mathrm{div}}(p,p)
=\frac{\alpha_s}{\pi}C_F\frac{1}{\epsilon}\gamma_5. 
\end{equation}
This is removed by Fig.~\ref{fig:pscala}(c), the ${\cal O}(\alpha_s)$ terms of
$Z_2$ and $Z^p$ being \cite{Larin:1993tq}
\begin{eqnarray}
Z_2&=&1-\frac{\alpha_s}{\pi}C_F\frac{1}{4\epsilon},
\nonumber\\
Z^p&=&1-\frac{\alpha_s}{\pi}C_F\frac{3}{4\epsilon}.
\end{eqnarray}
The remaining terms on the right-hand side of Eq.~\eqref{eq:pnum} are
evanescent; they live in the unphysical $(d-4)$-dimensional part of space-time
and vanish if we let $d\to 4$.
Yet the loop integral is divergent, so that an unphysical finite contribution
remains.
We can ensure the vanishing of such contributions to all orders by a finite
renormalisation.
This is exactly what is achieved by the finite renormalisation $Z_5^p$.
In this way, we ensure that the evanescent part does not mix into the physical
part.
Upon loop integration, the evanescent part of Eq.~(\ref{eq:pnum}) yields
\begin{equation}
\Gamma_5^{(1),\,\mathrm{eva}}(p,p)
=2\frac{\alpha_s}{\pi}C_F\gamma_5, 
\end{equation}
so that, through ${\cal O}(\alpha_s)$, the finite counterterm is
\begin{equation}
Z_5^p = 1+\delta Z_5^p=1 - 2\frac{\alpha_s}{\pi}C_F,
\label{eq:z5p}
\end{equation}
in agreement with Ref.~\cite{Larin:1993tq}. 

\begin{figure}[ht]
\begin{center}
\includegraphics[width=0.75\textwidth,viewport=110 668 428 730,clip]{%
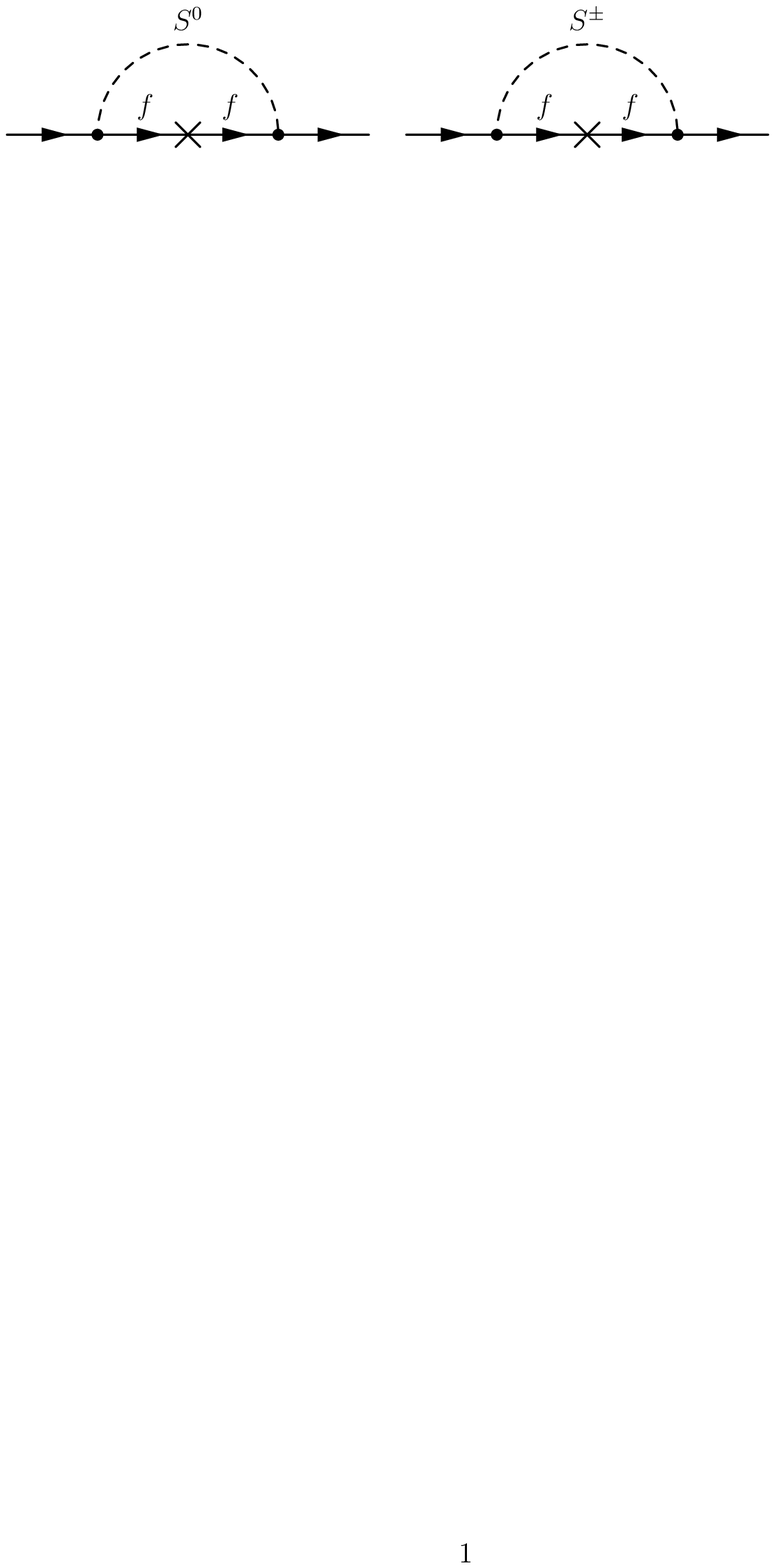}
\centerline{\hspace{4.6cm}(a)\hfill(b)\hspace{4.6cm}}
\caption{\label{fig:pscalaEW}%
Feynman diagrams contributing to $Z_5^p$ at ${\mathcal O}(x_t)$ and
${\mathcal O}(x_F)$.
Crosses indicate insertions of $P(x)$, and $S^0=\chi^0,h^0,H^0,A^0$,
$S^\pm=\phi^\pm,H^\pm$, and $f=t,b,U,D,N,E$ denote generic neutral and charged
scalar bosons and fermions, respectively.}
\end{center}
\end{figure}

We now apply the same procedure at ${\mathcal O}(x_t)$ and ${\mathcal O}(x_F)$.
To this end, we have to consider the counterparts of the diagram in
Fig.~\ref{fig:pscala}(b) where the gluon is replaced by neutral or charged
scalar electroweak bosons, $S^0=\chi^0,h^0,H^0,A^0$ and $S^\pm=\phi^\pm,H^\pm$,
where $\chi^0$ and $\phi^\pm$ denote the Goldstone bosons.
These are depicted in Figs.~\ref{fig:pscalaEW}(a) and (b), respectively.
We find that the sets of diagrams in Figs.~\ref{fig:pscalaEW}(a) and (b) add
up to zero separately.
In the three-generation case, the diagrams in Figs.~\ref{fig:pscalaEW}(b) do
not contribute in ${\mathcal O}(x_t)$ at all.
Consequently, we have $\delta Z_5^p=0$ at ${\mathcal O}(x_t)$ and
${\mathcal O}(x_F)$.


\boldmath
\section{\label{sec::pseudoscalar}$A^0\to\gamma\gamma$ and $A^0\to gg$}
\unboldmath

In this section, we first discuss the case of $A^0\to\gamma\gamma$ in detail.
Specifically, we recall the Born result and recover the
${\mathcal O}(\alpha_s)$ correction in Subsection~\ref{sec::born} and
evaluate the ${\mathcal O}(x_t)$ and ${\mathcal O}(x_F)$ corrections in
Subsection~\ref{sec::pseudoew}.
In Subsection~\ref{sec::agg}, we then extract from the latter the
corresponding corrections for $A^0\to gg$.

We drop all terms including an even number of $\gamma_5$ matrices in the
fermion trace, since they do not give contributions to the final results.
Those with an odd number lead to expressions that are proportional to the
epsilon tensor.
The transition amplitude can thus be decomposed as follows:
\begin{equation}\label{eq:A0lorentz}
\mathcal{T}=\frac{1}{4!}\varepsilon_{\alpha\beta\gamma\delta}
\epsilon_\mu^*(q_1)\epsilon_\nu^*(q_2)
\mathcal{A}^{\mu\nu\alpha\beta\gamma\delta},
\end{equation}
where $q_i$ and $\epsilon^*(q_i)$ are the four-momenta and polarisation
four-vectors of the outgoing photons $i=1,2$.
By Lorentz covariance, Eq.~(\ref{eq:A0lorentz}) can be written as 
\begin{equation}
\mathcal{T}=\varepsilon^{\mu\nu\rho\sigma}q_{1\mu}q_{2\nu}
\epsilon_\rho^*(q_1)\epsilon_\sigma^*(q_2)\mathcal{A},
\end{equation}
where
\begin{equation}
\mathcal{A}=-\frac{\tilde{g}_{[\alpha\mu}\tilde{g}_{\beta\nu}g_{\gamma\rho}
g_{\delta\sigma]_{1}}q_{1}^{\rho}q_{2}^{\sigma}}{48(q_1\cdot q_2)^{2}}
\mathcal{A}^{\mu\nu\alpha\beta\gamma\delta}.
\end{equation}
Here, $[\ ]_{1}$ denotes antisymmetrisation in the first indices of the metric
tensors, and it is understood that the external momenta $q_1$ and $q_2$ have
non-vanishing components only in the physical four dimensions of space-time.
The partial width of the $A^0\to\gamma\gamma$ decay is then given by
\begin{equation}
\Gamma(A^0\to\gamma\gamma)=\frac{M_{A^0}^{3}}{64\pi}|\mathcal{A}|^2.
\label{eq:app}
\end{equation}
The form factor $\mathcal{A}$ is evaluated in perturbation theory as 
\begin{equation}\label{eq:Notation}
\mathcal{A}=\sum_f\left(\mathcal{A}_f^{\mathrm{LO}}
+{\mathcal A}_f^{\alpha_s}+{\mathcal A}_f^{x_f}+\ldots\right)+\ldots,
\end{equation}
where the sum is over heavy fermions $f=t,F$, ${\mathcal A}_f^{\mathrm{LO}}$
is the one-loop contribution induced by charged fermions,
${\mathcal A}_f^{\alpha_s}$ is the two-loop QCD correction in the case of $f$
being a quark, ${\cal A}_f^{x_f}$ is the dominant two-loop electroweak
correction due to weak-isospin doublets of quarks and leptons, and the
ellipses stand for the residual one- and two-loop contributions as well as all
contributions beyond two loops.
The counterpart of Eq.~(\ref{eq:app}) for $\Gamma(A^0\to gg)$ contains an
additional colour factor of $N_g/4=2$ on the right-hand side.

The couplings of the $A^0$ boson to fermions are proportional to their masses.
Therefore, we only consider diagrams where the $A^0$ boson couples to the top
quark or a fourth-generation fermion.
However, we must bear in mind that diagrams where it couples to the bottom
quark may become sizeable for large values of $\tan\beta$, for
$\tan\beta={\mathcal O}(m_t/m_b)$, because that coupling is proportional to
$\tan\beta$.
In the following, we thus concentrate on low to intermediate values of
$\tan\beta$.


\boldmath
\subsection{\label{sec::born}Born result for $\Gamma(A^0\to\gamma\gamma)$ and
${\mathcal O}(\alpha_s)$ correction}
\unboldmath

\begin{figure}[ht]
\begin{center}
\includegraphics[width=0.75\textwidth,viewport=213 656 390 722,clip]{%
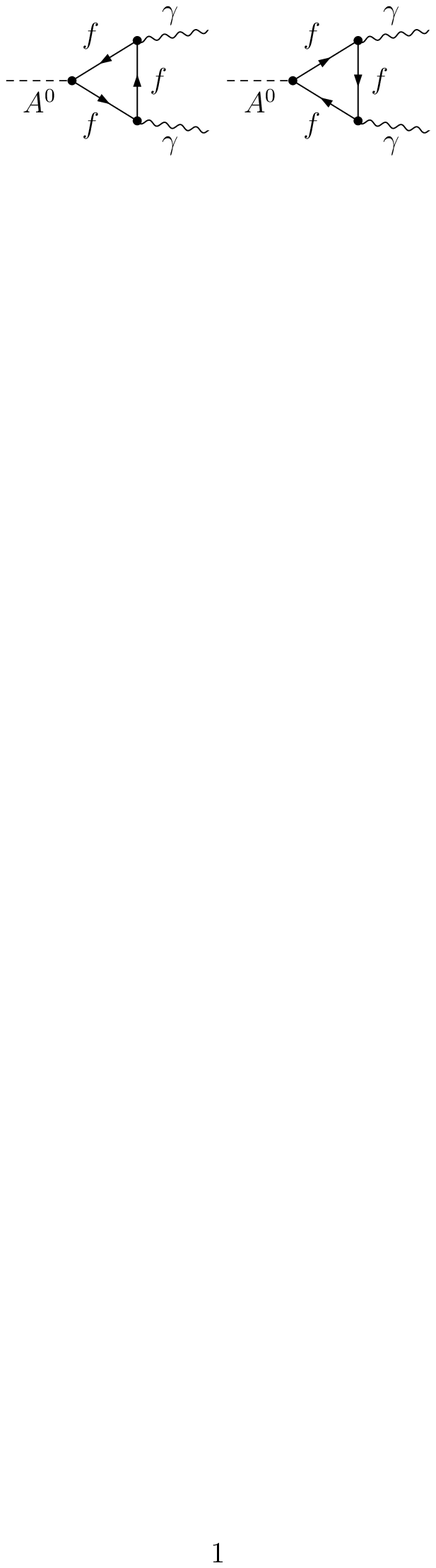}
\caption{\label{A01}
One loop diagrams contributing to $\Gamma(A^0 \to \gamma \gamma)$.
$f=t,b,U,D,N,E$ denotes generic fermions.}
\end{center}
\end{figure}
The LO result arises from the diagrams shown in Fig.~\ref{A01} and may be
found in Ref.~\cite{Kalyniak:1985ct} for $d=4$ space-time dimensions.
For $d=4-2\epsilon$ through ${\mathcal O}(\epsilon)$, we have in closed form
and as an expansion in $\tau_f=M_{A^0}^2/(2 m_f)^2$:
\begin{eqnarray}
{\mathcal A}_f^{\mathrm{LO}}&=&2^{1/4}G_F^{1/2}\frac{\alpha_{\mathrm{em}}}{\pi}
N_fQ_f^2g_f
\left(\frac{4\pi\mu^2}{m_f^2}\mathrm{e}^{-\gamma_E}\right)^\epsilon
\left[-\frac{1}{\tau_f}\arcsin^2{\sqrt{\tau_f}}+{\mathcal O}(\epsilon)\right]
\nonumber\\
&=&2^{1/4}G_F^{1/2}\frac{\alpha_{\mathrm{em}}}{\pi}N_fQ_f^2g_f
\left(\frac{4\pi\mu^2}{m_f^2}\mathrm{e}^{-\gamma_E}\right)^\epsilon
\left[-1-\frac{1}{3}\tau_f-\frac{8}{45}\tau_f^2 -\frac{4}{35}\tau_f^3
-\frac{128}{1575}\tau_f^4 +{\mathcal O}(\tau_f^5)\right.
\nonumber\\
&&{}+\left.\vphantom{\frac{1}{3}}
{\mathcal O}(\epsilon)\right],
\label{eq:res1lA0}
\end{eqnarray}
where $\gamma_E$ is the Euler-Mascheroni constant, $\alpha_{\mathrm{em}}$ is
Sommerfeld's fine-structure constant, $N_f=1,N_c$ is the colour multiplicity
of fermion $f$, $Q_f$ is its fractional electric charge, and
$g_f=\cot\beta,\tan\beta$ for up-type and down-type is its coupling strength
to the $A^0$ boson normalised to its Yukawa coupling in the SM.
Note that $M_{A^0}^2$ enters Eq.~(\ref{eq:res1lA0}) through the kinematic
relation $(q_1+q_2)^2=M_{A^0}^2$.
The corresponding result for $A^0\to gg$ is obtained from
Eq.~(\ref{eq:res1lA0}) through the substitution 
$\alpha_{\mathrm{em}}N_fQ_f^2\to\alpha_s$. 

\begin{figure}[ht]
\begin{center}
\includegraphics[width=0.95\textwidth,viewport=123 653 479 724,clip]{%
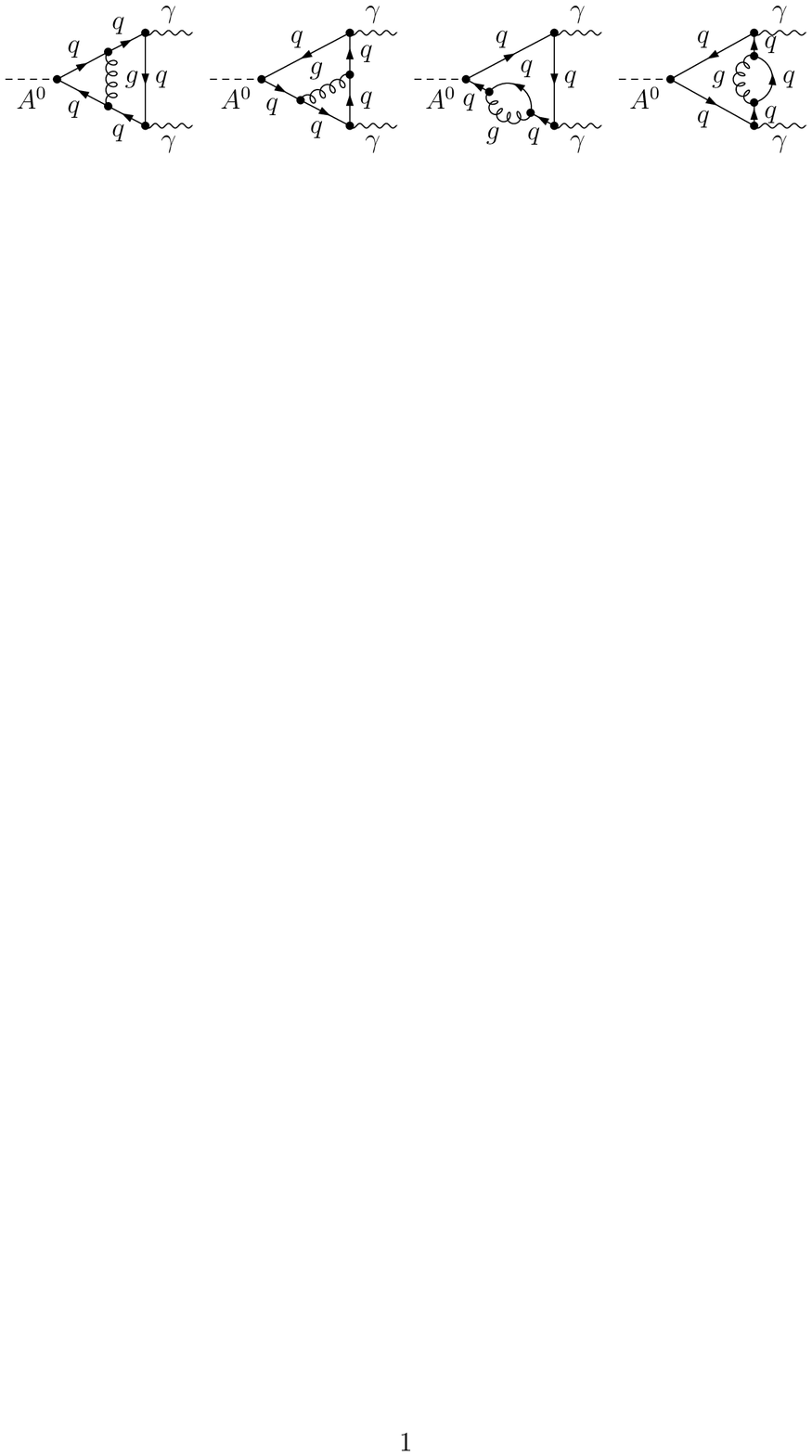}
\caption{\label{A0QCD}%
Sample two-loop diagrams contributing to the ${\mathcal O}(\alpha_s)$
QCD corrections to $\Gamma(A^0 \to \gamma \gamma)$.}
\end{center}
\end{figure}
We now turn to the ${\mathcal O}(\alpha_s)$ QCD corrections.
Besides the proper vertex diagrams, some of which are depicted in
Fig.~\ref{A0QCD}, we also need to include the counterterm contribution, so
that
${\mathcal A}_q^{\alpha_s}={\mathcal A}_{q,\mathrm{CT}}^{\alpha_s}
+{\mathcal A}_{q,0}^{\alpha_s}$.
As for renormalisation, besides including the finite counterterm of the
pseudoscalar current in Eq.~(\ref{eq:z5p}), we only need to renormalise the
quark mass appearing in the prefactor of Eq.~(\ref{eq:res1lA0}), by shifting
its bare value as $m_q^0=m_q+\delta m_q$, so that
\begin{equation}
{\mathcal A}_{q,\mathrm{CT}}^{\alpha_s}=
\left.{\mathcal A}_q^{\mathrm{LO}}\right|_{\epsilon=0}
\left(\delta Z_5^p-2\epsilon\frac{\delta m_q}{m_q}\right).
\label{eq:qcdct}
\end{equation}
In the on-shell scheme, the quark mass counterterm is
\begin{equation}
\frac{\delta m_q}{m_q}=\frac{\alpha_s}{\pi}C_F\left(-\frac{3}{4}\Delta
-\frac{3}{4}\ln\frac{\mu^2}{m_q^2}-1\right),
\end{equation}
where $\Delta=1/\epsilon-\gamma_E+\ln(4\pi)$.
In total, our evaluation yields
\begin{equation}\label{eq:res2laqcd}
{\cal A}_q^{\alpha_s}=
2^{1/4}G_F^{1/2}\frac{\alpha_{\mathrm{em}}}{\pi}\,\frac{\alpha_s}{\pi}
N_cQ_q^2g_q
\left[-\frac{16}{9}\tau_q-\frac{68}{45}\tau_q^2-\frac{53012}{42525}\tau_q^3
-\frac{34712}{33075}\tau_q^4+{\mathcal O}(\tau_q^5)\right],
\end{equation}
which agrees with the Taylor expansion of the analytic result derived in
Refs.~\cite{Harlander:2005rq,Aglietti:2006tp} from the integral representation
originally obtained in Refs.~\cite{Djouadi:1993ji,Spira:1995rr}.

Notice that the ${\mathcal O}(\tau^0)$ term in Eq.~(\ref{eq:res2laqcd})
vanishes, so that the ${\mathcal O}(\alpha_s)$ correction is suppressed for
small values of $M_{A^0}$.
In fact, as a consequence of the Adler-Bardeen theorem \cite{Adler:1969er},
the large-$m_t$ effective Lagrangian of the $A^0\gamma\gamma$ interaction does
not receive QCD corrections at any order
\cite{Djouadi:1993ji,Kauffman:1993nv,Kniehl:1995tn}.


\boldmath
\subsection{\label{sec::pseudoew}${\mathcal O}(x_t)$ and ${\mathcal O}(x_F)$
corrections to $\Gamma(A^0\to\gamma\gamma)$}
\unboldmath

\begin{figure}[ht]
\begin{center}
\includegraphics[width=0.95\textwidth,viewport=123 474 479 724,clip]{%
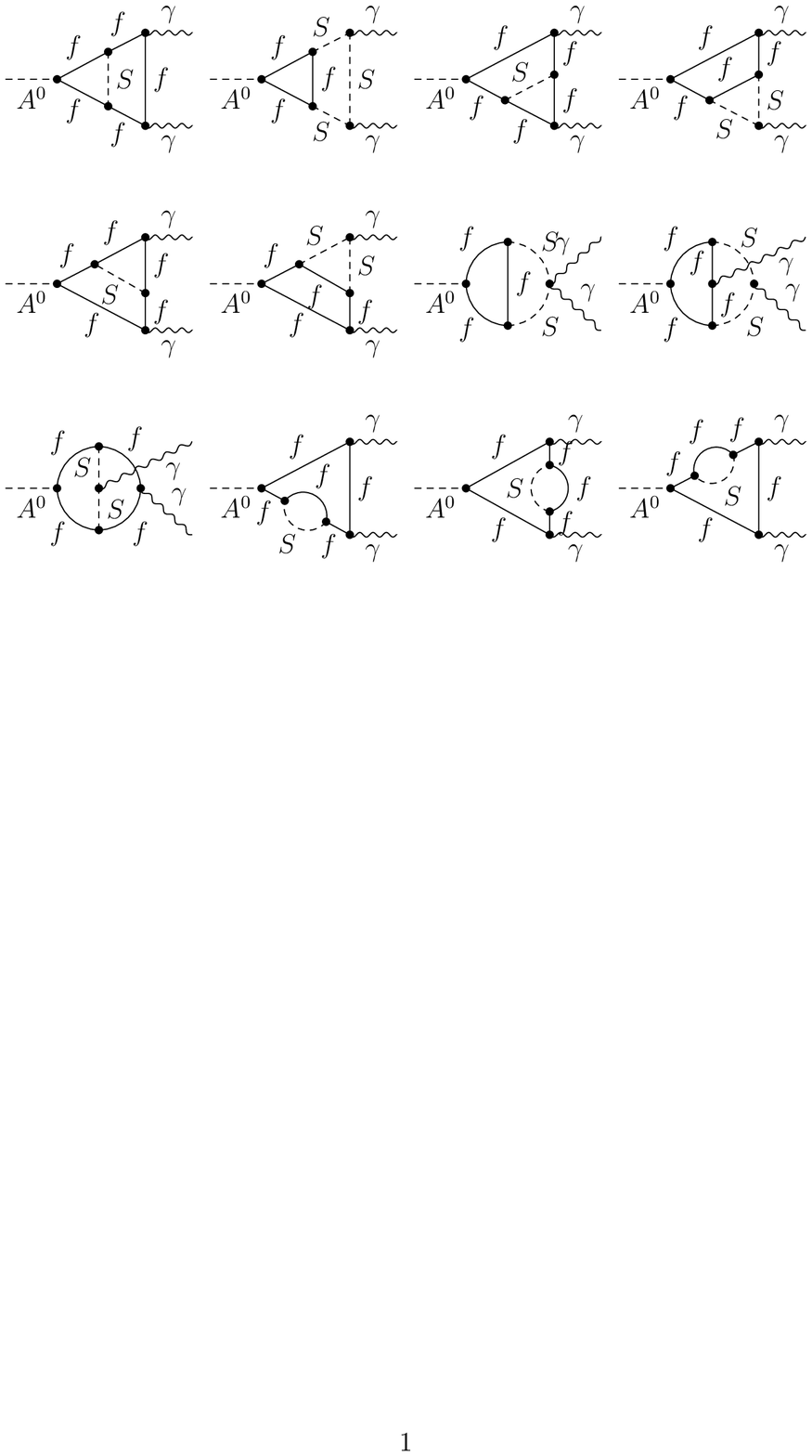}
\caption{\label{fi:ewa0}%
Generic two-loop diagrams contributing to the ${\mathcal O}(x_f)$
electroweak corrections to $\Gamma(A^0\to\gamma\gamma)$.
$S=\chi^0,\phi^\pm,h^0,H^0,A^0,H^\pm$ and $f=t,b,U,D,N,E$ denote generic
scalar bosons and fermions, respectively.
The couplings of the neutral scalar bosons to the bottom quark are to be
neglected and those of the neutral particles to the photon vanish.}
\end{center}
\end{figure}
Now we turn to the dominant electroweak two-loop corrections.
We first consider the case of three fermion generations.
Later on, we also study the additional corrections induced by a sequential
fermion generation, consisting of an up-type quark $U$, a down-type quark $D$,
a Dirac neutrino $N$, and a charged lepton $E$.
The contributing diagrams are depicted generically in Fig.~\ref{fi:ewa0}.
We have checked explicitly that diagrams including virtual $W$ bosons do not
contribute to our order.

Let us first discuss the counterterm contributions.
In contrast to the QCD case in Eq.~(\ref{eq:qcdct}), we now also have to
renormalise the $A^0$ wave function, $G_F$, and $\tan\beta$.
These additional contributions are universal.
We thus have
\begin{equation}
{\mathcal A}_{f,\mathrm{CT}}^{x_f}=
-2^{1/4}G_F^{1/2}\frac{\alpha_{\mathrm{em}}}{\pi}N_fQ_f^2g_f
\left(\delta Z_5^p-2\epsilon\frac{\delta m_f}{m_f}+\delta_u\right),
\label{eq:afct}
\end{equation}
where, in the electroweak on-shell scheme \cite{Sirlin:1980nh,Hollik:2002mv}
supplemented with the DCPR \cite{Chankowski:1992er} definition of $\tan\beta$,
\begin{equation}
\delta_u=\frac{\delta v}{v}-\frac{\Delta r}{2}.
\end{equation}
Here, $\delta v/v$ is the common DCPR counterterm for the two Higgs doublets
given in Eq.~(3.11) of Ref.~\cite{Dabelstein:1995js} and $\Delta r$
\cite{Sirlin:1980nh} contains those radiative corrections to the muon lifetime
which the SM introduces on top of those derived in the QED-improved Fermi
model.
In terms of (transverse) self-energies, we have
\begin{equation}
\delta_u=\frac{1}{2}
\left[-\frac{\Sigma_{W^\pm,T}(0)}{M_W^2}-\Sigma_{A^0}^\prime(M_{A^0}^2)
+(\tan\beta-\cot\beta)\frac{\Sigma_{A^0Z^0}(M_{A^0}^2)}{M_Z}\right].
\label{eq:delu}
\end{equation}

In the three-generation case, we set $m_b=0$ and formally impose the
following mass hierarchies:
\begin{equation}
M_Z,M_W,M_{h^0},M_{H^0},M_{A^0},M_{H^{\pm}}<m_t,\qquad
M_{A^0}<2M_W,2M_{H^{\pm}},
\label{eq:hierarch}
\end{equation}
which ensure the applicability of the asymptotic-expansion technique.
In practice, this implies that the unknown Higgs-boson masses obey
\begin{equation}
M_{h^0},M_{H^0},M_{A^0},M_{H^{\pm}}<m_t,\qquad 
M_{A^0}<2M_{H^{\pm}}.
\end{equation}
The leading two-loop electroweak corrections are then of ${\mathcal O}(x_t)$.
In the presence of fourth-generation fermions $F=U,D,N,E$, we assume that
their masses obey
\begin{equation}
M_Z,M_W,M_{h^0},M_{H^0},M_{A^0},M_{H^{\pm}}<m_U,m_D,m_N,m_E.
\end{equation}
For simplicity, we consider the special cases $m_U\gg m_D$, $m_U=m_D$, and
$m_U\ll m_D$, and similarly for the leptonic weak-isospin doublet, so that we
are effectively dealing with single-scale problems yielding corrections of
${\mathcal O}(x_F)$.

We first list the non-vanishing counterterms entering Eq.~(\ref{eq:afct}) for
a generic quark doublet $(U,D)$.
For the $U$ quark, we have
\begin{equation}
\frac{\delta m_U}{m_U}=\begin{cases}
\displaystyle\frac{x_U}{\sin^2\beta}\left(\frac{3}{2}\Delta
+\frac{3}{2}\ln\frac{\mu^2}{m_U^2}+4\right) &
\text{if $m_U\gg m_D$},\\
\displaystyle x_U\left(\frac{3}{\sin^2\beta}+\frac{1}{\cos^2\beta}\right)
\left(\frac{1}{2}\Delta
+\frac{1}{2}\ln\frac{\mu^2}{m_U^2}+\frac{3}{2}\right) &
\text{if $m_U=m_D$},\\
\displaystyle\frac{x_D}{\cos^2\beta}\left(\frac{1}{2}\Delta
+\frac{1}{2}\ln\frac{\mu^2}{m_D^2}+\frac{3}{4}\right) &
\text{if $m_U\ll m_D$}.
\end{cases}
\label{eq:cAA}
\end{equation}
The corresponding expression for the $D$ quark is obtained from
Eq.~(\ref{eq:cAA}) by interchanging $m_U\leftrightarrow m_D$ and
$\sin\beta\leftrightarrow\cos\beta$.
Furthermore, we have
\begin{equation}
\delta_u=\begin{cases}
\displaystyle\frac{N_c}{2}x_U & \text{if $m_U\gg m_D$,}\\
0& \text{if $m_U=m_D$,}\\
\displaystyle\frac{N_c}{2}x_D & \text{if $m_U\ll m_D$.}
\end{cases}
\label{eq:deltau}
\end{equation}
The counterparts of Eqs.~(\ref{eq:cAA}) and (\ref{eq:deltau}) for a generic
lepton doublet $(N,E)$ are obtained by substituting $m_U\to m_N$,
$m_D\to m_E$, and $N_c\to1$.

Our final result for the three-generation case reads
\begin{eqnarray}
{\mathcal A}_t^{x_t}&=&2^{1/4}G_F^{1/2}\frac{\alpha_{\mathrm{em}}}{\pi}
N_cx_t\cot\beta
\left(\frac{20}{9}
+\frac{6}{9}
-\frac{12}{9}\frac{\cos^2{\alpha}}{\sin^2{\beta}}
-\frac{12}{9}\frac{\sin^2{\alpha}}{\sin^2{\beta}}
+\frac{20}{9}\cot^2{\beta}
+\frac{6}{9}\cot^2{\beta}
\right.
\nonumber\\
&&{}-\left.\frac{2}{9}N_c\right)
\label{eq:astandard}\\
&=&2^{1/4}G_F^{1/2}\frac{\alpha_{\mathrm{em}}}{\pi}N_cx_t
\left[\frac{14}{9}\cot^3\beta+\frac{2}{9}(7-N_c)\cot\beta\right],
\end{eqnarray}
where $\alpha$ is the angle that rotates the weak eigenstates of the neutral
CP-even Higgs bosons into their mass eigenstates.
On the right-hand side of Eq.~(\ref{eq:astandard}), we exhibit separately the
finite contributions from the $\chi^0$, $\phi^\pm$, $h^0$, $H^0$, $A^0$, and
$H^\pm$ bosons, and the universal counterterm, after top-quark mass
renormalisation, in the order in which they appear there.
The UV-divergent parts vanish after exploiting simple trigonometric
identities.
The same is true for the unrenormalised contributions from neutral particles
as well as for the na\"\i ve contributions from the asymptotic expansion of
the diagrams containing charged particles.
However, the total result is nonzero, as opposed to the QCD case.
We observe that the $\alpha$ dependence carried by the contributions from the
neutral CP-even Higgs bosons cancels in their sum.
This reflects the fact that, by neglecting their masses, we effectively treat
the $h^0$ and $H^0$ bosons as mass degenerate, so that we may rotate the
angle $\alpha$ away.

Now we examine the influence of a sequential generation consisting of heavy
fermions $F$ as specified above.
Besides the appearance of new leading correction terms quadratic in their
masses, of generic order ${\mathcal O}(x_F)$, also the ${\mathcal O}(x_t)$
correction is then modified.
This may be understood by observing that the LO result then receives three
more mass-independent contributions in addition to the one from the top quark,
from the charged fermions $U,D,E$, which feed into the ${\mathcal O}(x_t)$
correction through the universal counterterm $\delta_u$.
This may be accommodated in Eq.~(\ref{eq:astandard}) through the substitution
\begin{equation}
-\frac{2}{9}N_c\to-\frac{2}{9}N_c-\frac{2}{9}N_c-\frac{N_c}{18}\tan^2\beta
-\frac{1}{2}\tan^2\beta.
\end{equation}

The ${\mathcal O}(x_F)$ contribution due to the $(U,D)$ doublet is found to be
\begin{equation}
\frac{{\mathcal A}_{(U,D)}^{x_F}}{2^{1/4}G_F^{1/2}(\alpha_{\mathrm{em}}/\pi)
N_c}=
\begin{cases}
\displaystyle x_U\left(\frac{4}{3}\cot^3\beta+\frac{13-4N_c}{9}\cot\beta
-\frac{7+N_c}{18}\tan\beta\right) & \text{if $m_U\gg m_D$,}\\
\displaystyle x_U\left(\frac{4}{3}\cot^3\beta+\frac{17}{9}\cot\beta
+\frac{8}{9}\tan\beta+\frac{1}{3}\tan^3\beta\right) & \text{if $m_U=m_D$,}\\
\displaystyle x_D\left[\frac{4}{9}(1-N_c)\cot\beta+\frac{5-N_c}{18}\tan\beta 
+\frac{1}{3}\tan^3\beta\right] & \text{if $m_U\ll m_D$.}
\end{cases}
\label{eq:UD}
\end{equation}
In each case, we find that the contributions from the various proper two-loop
diagrams cancel, so that we are only left with the counterterm contributions.
Again, the total result is non-zero, in contrast to the QCD case.

The counterpart of Eq.~(\ref{eq:UD}) for the $(N,E)$ doublet is simply
obtained by appropriately adjusting the quantum numbers $N_f$ and $Q_f$ and
reads:
\begin{equation}
\frac{{\mathcal A}_{(N,E)}^{x_F}}{2^{1/4}G_F^{1/2}(\alpha_{\mathrm{em}}/\pi)}=
\begin{cases}
\displaystyle x_N\left[\left(1-\frac{4}{9}N_c\right)\cot\beta
+\frac{1}{2}\left(1-\frac{N_c}{9}\right)\tan\beta
\right ]& \text{if $m_N\gg m_E$,}\\
\displaystyle x_N\left(\cot\beta+4\tan\beta+3\tan^3\beta\right) &
\text{if $m_N=m_E$,}\\
\displaystyle x_E\left[-\frac{4}{9}N_c\cot\beta
+\frac{1}{2}\left(5-\frac{N_c}{9}\right)\tan\beta+3\tan^3\beta\right] &
\text{if $m_N\ll m_E$.}
\end{cases}
\label{eq:NE}
\end{equation}

\boldmath
\subsection{\label{sec::agg}${\mathcal O}(x_t)$ and ${\mathcal O}(x_F)$
corrections to $\Gamma(A^0\to gg)$}
\unboldmath

Now we turn to the ${\mathcal O}(x_t)$ and ${\mathcal O}(x_F)$ corrections to
$\Gamma(A^0\to gg)$.
They may be easily extracted from the analogous calculation for
$\Gamma(A^0\to\gamma\gamma)$ discussed in Section~\ref{sec::pseudoew}, by
retaining only those diagrams where both photons couple to quark lines and
substituting $\alpha_{\mathrm{em}}Q_qQ_{q^\prime}N_c\to\alpha_s$.
Note that this does not affect the factors of $N_c$ originating from the
renormalisation procedure.

In the three-generation case, we thus obtain
\begin{eqnarray}
{\cal A}_t^{x_t}&=&2^{1/4}G_F\frac{\alpha_s}{\pi}x_t\cot\beta
\left(5
+3
-3\frac{\cos^2\alpha}{\sin^2\beta}
-3\frac{\sin^2\alpha}{\sin^2\beta}
+5\cot^2\beta
+3\cot^2\beta
-\frac{N_c}{2}\right)
\label{eq:aggsm}\\
&=&2^{1/4}G_F\frac{\alpha_s}{\pi}x_t
\left[5\cot^3\beta+\left(5-\frac{N_c}{2}\right)\cot\beta\right],
\end{eqnarray}
where the five terms on the right-hand side of Eq.~(\ref{eq:aggsm}) again
represent the finite contributions from the $\chi^0$, $\phi^\pm$, $h^0$,
$H^0$, $A^0$, and $H^\pm$ bosons, and the universal counterterm, after
top-quark mass renormalisation.
The final result is again independent of $\alpha$.

In the presence of a sequential heavy-fermion generation, the universal
counterterm in Eq.~(\ref{eq:aggsm}) is modified according to
\begin{equation}
-\frac{N_c}{2}\to-\frac{N_c}{2}-\frac{N_c}{2}-\frac{N_c}{2}\tan^2\beta.
\end{equation}

The ${\mathcal O}(x_F)$ contribution due to the $(U,D)$ doublet is found to be
\begin{equation}
\frac{{\mathcal A}_{(U,D)}^{x_F}}{2^{1/4}G_F^{1/2}(\alpha_s/\pi)}=
\begin{cases}
\displaystyle x_U\left[3\cot^3\beta+(4-N_c)\cot\beta
+\left(1-\frac{N_c}{2}\right)\tan\beta\right] & \text{if $m_U\gg m_D$,}\\
\displaystyle x_U(3\cot^3\beta+5\cot\beta+5\tan\beta+3\tan^3\beta) &
\text{if $m_U=m_D$,}\\
\displaystyle x_D\left[(1-N_c)\cot\beta+\left(4-\frac{N_c}{2}\right)\tan\beta 
+3\tan^3\beta\right] & \text{if $m_U\ll m_D$.}
\end{cases}
\label{eq:gUD}
\end{equation}

The $(N,E)$ doublet can generate a ${\mathcal O}(x_F)$ contribution only
through the universal counterterm, so that
\begin{equation}
\frac{{\mathcal A}_{(N,E)}^{x_F}}{2^{1/4}G_F^{1/2}(\alpha_s/\pi)}=
\begin{cases}
\displaystyle x_N\left(-\cot\beta-\frac{1}{2}\tan\beta\right) &
\text{if $m_N\gg m_E$,}\\
0 & \text{if $m_N=m_E$,}\\
\displaystyle x_E\left(-\cot\beta-\frac{1}{2}\tan\beta\right) &
\text{if $m_N\ll m_E$.}
\end{cases}
\label{eq:gNE}
\end{equation}

Note that the ${\mathcal O}(\alpha_s)$ correction to $\Gamma(A^0\to gg)$
cannot be recovered from the one to $\Gamma(A^0\to\gamma\gamma)$ because it
receives additional contributions from diagrams involving gluon self-couplings.

\boldmath
\section{\label{sec::scalar}${\mathcal O}(x_f)$ correction to
$\Gamma(H\to\gamma\gamma)$}
\unboldmath

Applying similar techniques as in Section~\ref{sec::pseudoew}, we now also
derive the ${\mathcal O}(x_f)$ correction to $\Gamma(H\to\gamma\gamma)$ in the
SM endowed with a sequential generation of heavy fermions.
Due to electromagnetic gauge invariance, the transition-matrix element of
$H\to\gamma\gamma$ possesses the structure
\begin{equation}
{\mathcal T}=[(q_1\cdot q_2)g^{\mu\nu}-q_1^{\nu}q_2^{\mu}]
\epsilon_\mu^*(q_1)\epsilon_\nu^*(q_2){\mathcal A}.
\label{eq:ahgg}
\end{equation}
To obtain a strong check on our analysis, we actually verify electromagnetic
gauge invariance by separately projecting out the coefficients of the Lorentz
tensors $(q_1\cdot q_2)g^{\mu\nu}$ and $q_1^{\nu}q_2^{\mu}$ in
Eq.~(\ref{eq:ahgg}).
Furthermore,
we work in
general $R_{\xi}$ gauge, so as to verify that the gauge parameter $\xi$
cancels in the final result.
From Eq.~(\ref{eq:ahgg}), we obtain
\begin{equation}
\Gamma(H\to\gamma\gamma)=\frac{M_H^{3}}{64\pi}|{\mathcal A}|^2,
\end{equation}
where $M_H$ is the mass of the SM Higgs boson.

The form factor ${\mathcal A}$ is evaluated in perturbation theory as 
\begin{equation}\label{eq:Notationa}
{\mathcal A}={\mathcal A}_W^{\mathrm{LO}}+\sum_f
\left({\mathcal A}_f^{\mathrm{LO}}+{\mathcal A}_f^{\alpha_s}
+{\mathcal A}_f^{x_f}+\cdots\right)+\cdots,
\end{equation}
where ${\cal A}_W^{\mathrm{LO}}$ denotes the one-loop contribution due to the
$W^\pm$ boson and the other contributions carry similar meanings as in
Eq.~(\ref{eq:Notation}).

A comprehensive review of the present theoretical knowledge of
$\Gamma(H\to\gamma\gamma)$ may be found in Ref.~\cite{Kniehl:1993ay}.
The LO result was first obtained in Ref.~\cite{Ellis:1975ap}.
The ${\mathcal O}(\alpha_s)$ \cite{Harlander:2005rq,Djouadi:1990aj} and
${\mathcal O}(\alpha_s^2)$ \cite{Steinhauser:1996wy} QCD corrections are
also available.
As for the two-loop electroweak correction, the contributions induced by light
\cite{Aglietti:2004nj} and heavy fermions \cite{Djouadi:1997rj,Fugel:2004ug}
as well as the residual ones \cite{Degrassi:2005mc} were recently evaluated.
The ${\mathcal O}(x_F)$ correction due to a sequential generation of heavy
fermions was studied in Ref.~\cite{Djouadi:1997rj} for general values of
their masses.
In the following, we revisit this analysis for the mass hierarchies
$m_U\gg m_D$, $m_U=m_D$, and $m_U\ll m_D$, and similarly for the leptons $N$
and $E$ using asymptotic expansion.
We refrain from considering $\Gamma(H\to gg)$.

\begin{figure}[ht]
\begin{center}
\includegraphics[width=0.95\textwidth,viewport=122 498 480 729,clip]{%
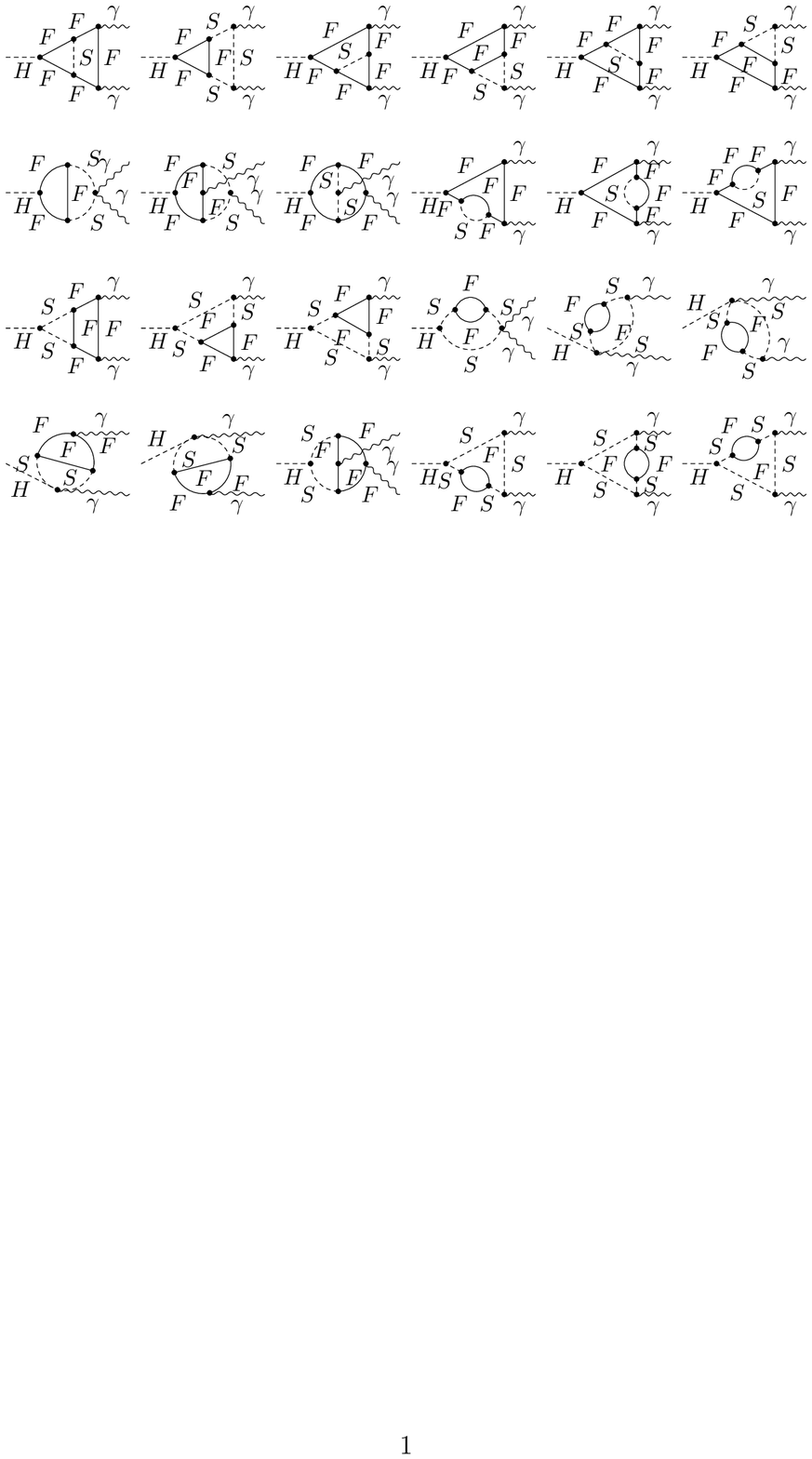}
\caption{\label{EW}%
Generic two-loop diagrams contributing to the ${\mathcal O}(x_F)$
electroweak correction to $\Gamma(H\to\gamma\gamma)$.
$S=\chi^0,\phi^\pm,W^\pm,H$ and $F=U,D,N,E$ denote generic bosons and
fermions, respectively.
The couplings of the neutral particles to the photon vanish.}
\end{center}
\end{figure}
The diagrams contributing to $\Gamma(H\to\gamma\gamma)$ at ${\mathcal O}(x_F)$
are shown generically in Fig.~\ref{EW}.
In contrast to the case of $A^0\to\gamma\gamma$, now also virtual $W^\pm$
bosons participate in ${\mathcal O}(x_F)$ \cite{Djouadi:1997rj,Fugel:2004ug}.

\begin{figure}[ht]
\begin{center}
\includegraphics[width=0.75\textwidth,viewport=154 539 450 725,clip]{%
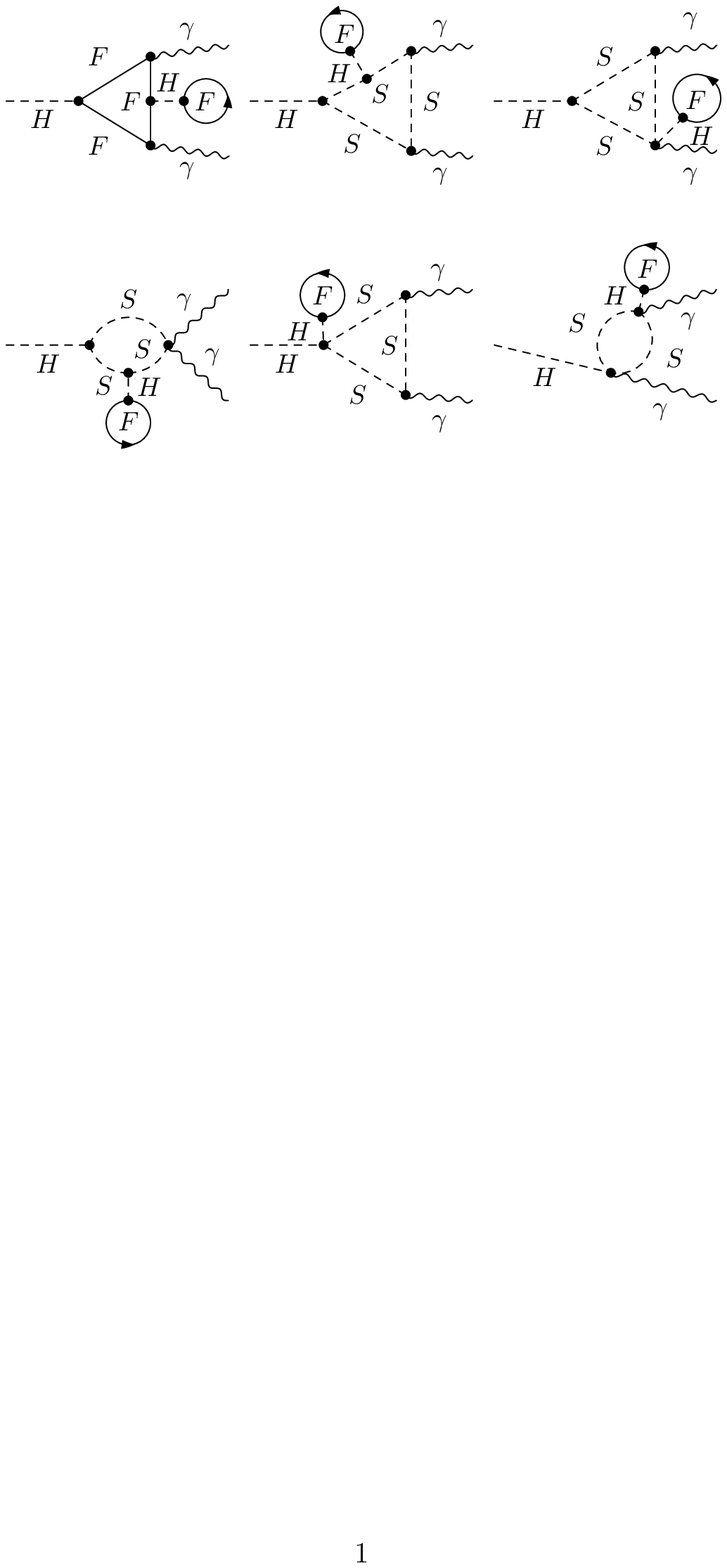}
\caption{\label{fi:tadquark}%
Examples of two-loop tadpole diagrams contributing to the ${\mathcal O}(x_F)$
electroweak correction to $\Gamma(H\to\gamma\gamma)$.
$S=\chi^0,\phi^\pm,W^\pm,H$ and $F=U,D,N,E$ denote generic bosons and
fermions, respectively.
The couplings of the neutral particles to the photon vanish.}
\end{center}
\end{figure}
Unlike in Refs.~\cite{Butenschoen:2006ns,Butenschoen:2007hz}, we choose not to
perform tadpole renormalisation here.
In turn, we need to include the diagrams that are generated by attaching a
Higgs tadpole in all possible ways to any one-loop seed diagram.
Some examples are depicted in Fig.~\ref{fi:tadquark}.
These tadpole diagrams yield terms proportional to $m_F^4$, which cancel
against the tadpole contributions to the counterterms, which are all
proportional to $m_F^4$, and the $m_F^4$ terms from the Higgs mass
renormalisation and the asymptotic expansion of the two-loop diagrams
involving virtual $\phi^\pm$ bosons.
Thus, the final result is devoid of $m_F^4$ terms.

It is interesting to note that the contribution from the attachment of a
tadpole to a fermion line is cancelled by the $m_F^4$ term from the
renormalisation of the mass of that line in the one-loop seed diagram.
Furthermore, the tadpole contributions to the renormalisations of the factors
$1/M_W$ and $m_F$ in the $HF\overline{F}$ vertex cancel each other.
These are the only tadpole contributions that could generate $m_F^2$ terms in
the final result, through the expansion in $\tau_F$.

Prior to evaluating the proper diagrams of Figs.~\ref{EW} and
\ref{fi:tadquark}, we discuss the renormalisation in some detail.
As before, we need to renormalise the Higgs-boson wave function and the masses
of the $W^\pm$ boson and the heavy fermions.
In addition, we now also need to renormalise the Higgs-boson mass.
The parameter $M_W$ appears in the $W^\pm$, $\phi^\pm$, and $u^\pm$
propagators and in the $HW^{\pm}W^{\mp}$, $H\phi^{\pm}W^{\mp}$,
$\phi^{\pm}W^{\mp}\gamma$, $H\phi^{\pm}W^{\mp}\gamma$, and $HF\overline{F}$
vertices, where $u^\pm$ are the charged Faddeev-Popov ghosts. 
The only vertex involving $M_H$ is $H\phi^{\pm}\phi^{\mp}$, which induces
two-loop contributions via $\delta M_H$.
Finally, $m_F$ occurs in the $F$-fermion propagator and in the
$HF\overline{F}$ vertex.
The corresponding counterterms are defined through
\begin{eqnarray}
  m_F^0 &=& m_F + \delta m_F + \delta m_F^{\mathrm{tad}} ,
  \nonumber \\
  (M_W^0)^2 &=& M_W^2 + \delta M_W^2 + \delta M_W^{2,\mathrm{tad}} ,
  \nonumber \\
  (M_H^0)^2 &=& M_H^2 + \delta M_H^2 + \delta M_H^{2,\mathrm{tad}} ,
  \nonumber \\
  H^0 &=& \sqrt{Z_H} H = \left(1+\frac{1}{2} \delta Z_H\right) H ,
\label{eq:smct}
\end{eqnarray}
where tadpole contributions are marked by the superscript ``tad.'' 
Note that $\delta Z_H$ is obtained from the derivative of the Higgs-boson
self-energy and thus has no tadpole contribution.

In the case of the $(U,D)$ doublet, the counterterms in Eq.~(\ref{eq:smct})
read:
\begin{eqnarray}
\frac{\delta m_U}{m_U}&=&\begin{cases}
\displaystyle x_U\left(\frac{3}{2}\Delta+\frac{3}{2}\ln\frac{\mu^2}{m_U^2}+4
\right)
& \text{if $m_U\gg m_D$},\\
2x_U
& \text{if $m_U=m_D$},\\
\displaystyle x_D\left(-\frac{3}{2}\Delta-\frac{3}{2}\ln\frac{\mu^2}{m_D^2}
-\frac{5}{4}\right)
& \text{if $m_U\ll m_D$},
\end{cases}
\nonumber\\
\frac{\delta m_U^{\mathrm{tad}}}{m_U}&=&\begin{cases}
\displaystyle x_U N_c\frac{m_U^2}{M_H^2}
\left(4\Delta+4\ln\frac{\mu^2}{m_U^2}+4\right)
& \text{if $m_U\gg m_D$},\\
\displaystyle x_U N_c\frac{m_U^2}{M_H^2}
\left(8\Delta+8\ln\frac{\mu^2}{m_U^2}+8\right)
& \text{if $m_U=m_D$},\\
\displaystyle x_D N_c\frac{m_D^2}{M_H^2}
\left(4\Delta+4\ln\frac{\mu^2}{m_D^2}+4\right)
& \text{if $m_U\ll m_D$},
\end{cases}
\nonumber\\
\frac{\delta M_W^2}{M_W^2}&=&\begin{cases}
\displaystyle x_U N_c\left(-2\Delta-2\ln\frac{\mu^2}{m_U^2}-1\right)
& \text{if $m_U\gg m_D$},\\
\displaystyle x_U N_c\left(-4\Delta-4\ln\frac{\mu^2}{m_U^2}\right)
& \text{if $m_U=m_D$},\\
\displaystyle x_D N_c\left(-2\Delta-2\ln\frac{\mu^2}{m_D^2}-1\right)
& \text{if $m_U\ll m_D$},
\end{cases}
\nonumber\\
\frac{\delta M_W^{2,{\mathrm{tad}}}}{M_W^2}&=&
2\frac{\delta m_U^{\mathrm{tad}}}{m_U},
\nonumber\\
\frac{\delta M_H^2}{M_H^2}&=&\begin{cases}
\displaystyle x_U N_c\left[\frac{m_U^2}{M_H^2}
\left(-12\Delta-12\ln\frac{\mu^2}{m_U^2}-4\right)
+2\Delta+2\ln\frac{\mu^2}{m_U^2}-\frac{4}{3}\right]
& \text{if $m_U\gg m_D$},\\
\displaystyle x_U N_c\left[\frac{m_U^2}{M_H^2}
\left(-24\Delta-24\ln\frac{\mu^2}{m_U^2}-8\right)
+4\Delta+4\ln\frac{\mu^2}{m_U^2}-\frac{8}{3}\right]
& \text{if $m_U=m_D$},\\
\displaystyle x_D N_c\left[\frac{m_D^2}{M_H^2}
\left(-12\Delta-12\ln\frac{\mu^2}{m_D^2}-4\right)
+2\Delta+2\ln\frac{\mu^2}{m_D^2}-\frac{4}{3}\right]
& \text{if $m_U\ll m_D$},
\end{cases}
\nonumber\\
\frac{\delta M_H^{2,{\mathrm{tad}}}}{M_H^2}&=&
3\frac{\delta m_U^{\mathrm{tad}}}{m_U},
\nonumber
\end{eqnarray}

\begin{eqnarray}
\delta Z_H&=&\begin{cases}
\displaystyle x_U N_c\left(-2\Delta-2\ln\frac{\mu^2}{m_U^2}+\frac{4}{3}\right)
& \text{if $m_U\gg m_D$},\\
\displaystyle x_U N_c\left(-4\Delta-4\ln\frac{\mu^2}{m_U^2}+\frac{8}{3}\right)
& \text{if $m_U=m_D$},\\
\displaystyle x_D N_c\left(-2\Delta-2\ln\frac{\mu^2}{m_D^2}+\frac{4}{3}\right)
& \text{if $m_U\ll m_D$},
\end{cases}
\label{eq:smcts}
\end{eqnarray}
and similarly for $\delta m_D/m_D$ and $\delta m_D^{\mathrm{tad}}/m_D$.
The renormalisations of the Higgs-boson wave function and the $W$-boson mass
in the $HF\overline{F}$ Yukawa coupling combine to a universal correction
\cite{Kniehl:1993jc},
\begin{equation}
\delta_u=\frac{1}{2}\left(\delta Z_H-\frac{\delta M_W^2}{M_W^2}\right),
\end{equation}
which should be compared with Eq.~(\ref{eq:delu}) for the case of the $A^0$
boson.
We have \cite{Chanowitz:1978uj}
\begin{equation}
\delta_u=\begin{cases}
\displaystyle\frac{7}{6}x_U N_c & \text{if $m_U\gg m_D$,}
\vphantom{\bigg|}
\\
\displaystyle\frac{4}{3}x_U N_c & \text{if $m_U=m_D$,}
\vphantom{\bigg|}
\\
\displaystyle\frac{7}{6}x_D N_c & \text{if $m_U\ll m_D$.}
\vphantom{\bigg|}
\end{cases}
\label{RenUniv}
\end{equation}
The counterparts of Eqs.~(\ref{eq:smcts}) and (\ref{RenUniv}) for the $(N,E)$
doublet are obtained by substituting $m_U\to m_N$, $m_D\to m_E$, and
$N_c\to1$.
Those for the three-generation SM may be found in Eqs.~(68)--(71), (74)--(76),
and (80) of Ref.~\cite{Butenschoen:2007hz}.

We now list our final results for ${\mathcal A}_f^{x_f}$ due to the $(U,D)$
doublet.
For the sake of comparison with Ref.~\cite{Djouadi:1997rj}, we exhibit the
dependence on the electric charge of the heavier quark, exploiting the
relation $Q_U=Q_D+1$.
We find
\begin{eqnarray}
\frac{{\mathcal A}_{(U,D)}^{x_F}}{2^{1/4}G_F^{1/2}(\alpha_{\mathrm{em}}/\pi)
N_c}&=&
\begin{cases}
\displaystyle x_U\left[
-\frac{25}{36}-6Q_U+4Q_U^2-\frac{7}{9}N_c\left(1-2Q_U+3Q_U^2\right)\right]
& \text{if $m_U\gg m_D$,}\\
\displaystyle x_U\left[
-\frac{56}{9}-8Q_U+8Q_U^2-\frac{8}{9}N_c\left(1-2Q_U+3Q_U^2\right)\right]
& \text{if $m_U=m_D$,}\\
\displaystyle x_D\left[
-\frac{25}{36}+6Q_D+4Q_D^2-\frac{7}{9}N_c\left(2+4Q_D+3Q_D^2\right)\right]
& \text{if $m_U\ll m_D$,}
\end{cases}
\nonumber\\
&=&
\begin{cases}
\displaystyle x_U\left(-\frac{35}{12}-\frac{7}{9}N_c\right)
& \text{if $m_U\gg m_D$,}\\
\displaystyle x_U\left(-8-\frac{8}{9}N_c\right)
& \text{if $m_U=m_D$,}\\
\displaystyle x_D\left(-\frac{9}{4}-\frac{7}{9}N_c\right)
& \text{if $m_U\ll m_D$.}
\end{cases}
\label{eq:smUD}
\end{eqnarray}

Appropriately adjusting the quantum numbers in the various contributions, we
find ${\mathcal A}_f^{x_f}$ due to the $(N,E)$ doublet to be
\begin{equation}
\frac{{\mathcal A}_{(N,E)}^{x_F}}{2^{1/4}G_F^{1/2}(\alpha_{\mathrm{em}}/\pi)}
=\begin{cases}
\displaystyle x_N\left(-\frac{25}{36}-\frac{7}{9}N_c\right)
& \text{if $m_N\gg m_E$,}\\
\displaystyle x_N\left(-\frac{56}{9}-\frac{8}{9}N_c\right)
& \text{if $m_N=m_E$,}\\
\displaystyle x_E\left(-\frac{97}{36}-\frac{7}{9}N_c\right)
& \text{if $m_N\ll m_E$.}
\end{cases}
\end{equation}

In the remainder of this section, we compare our results with those obtained
in Ref.~\cite{Djouadi:1997rj}.
That reference is more general than ours in the sense that no hierarchies
among the heavy-fermion masses are assumed.
However, we can compare Eq.~(\ref{eq:smUD}) with Eqs.~(56) and (57) of
Ref.~\cite{Djouadi:1997rj}, where the hierarchies $m_U\gg m_D$ and
$m_U=m_D$, respectively, are considered.
We can reproduce these equations if we include an overall minus sign in our
expression for $\delta m_U/m_U$ in Eq.~(\ref{eq:smcts}).
In other words, there should be an overall minus sign on the right-hand side
of Eq.~(9) in Ref.~\cite{Djouadi:1997rj}, which was already noticed in
Ref.~\cite{Fugel:2004ug} in connection with the SM case.
In Ref.~\cite{Fugel:2004ug}, it was also observed that the limit $m_D\to 0$ of
the fourth-generation result for $M_W\ll m_D\ll m_U$ differs from the
calculation with $m_D=0$ from the beginning, which is appropriate for the
third generation, where $m_b\ll M_W\ll m_t$.
These two observations lead to a modification of Eq.~(60) in
Ref.~\cite{Djouadi:1997rj}, where the leading correction due to the five heavy
fermions $t$, $U$, $D$, $N$, and $E$ is presented.
Adopting the notation of Ref.~\cite{Djouadi:1997rj}, we have
\begin{eqnarray}
A_{\mathrm{4gen}}&=&\frac{\alpha G_{\mu}^{1/2}}{\pi 2^{3/4}}\,\frac{5}{3}
\left[1+\frac{G_{\mu}}{8\pi^2\sqrt{2}}\left(-\frac{197}{10}m_t^2
-\frac{109}{30}m_N^2-\frac{181}{30}m_E^2
-\frac{m_N^2m_E^2}{m_N^2-m_E^2}\ln\frac{m_N^2}{m_E^2}\right.\right.
\nonumber\\
&&{}-\left.\left.
\frac{189}{10}m_U^2-\frac{33}{2}m_D^2
-3\frac{m_U^2m_D^2}{m_U^2-m_D^2}\ln\frac{m_U^2}{m_D^2}\right)\right],
\end{eqnarray}
where $\alpha=\alpha_{\mathrm{em}}$ and $G_\mu=G_F$.
By the same token, Eq.~(59) of Ref.~\cite{Djouadi:1997rj} becomes
\cite{Fugel:2004ug,Degrassi:2005mc}:
\begin{equation}
A_{\mathrm{SM}}=\frac{\alpha G_{\mu}^{1/2}}{\pi 2^{3/4}}\,\frac{47}{9}
\left[1+\frac{G_{\mu}}{8\pi^2\sqrt{2}}\left(-\frac{367}{94}m_t^2\right)
\right].
\end{equation}

In Ref.~\cite{Djouadi:1997rj}, the ${\mathcal O}(x_f)$ corrections to the
$Hgg$ coupling in the SM with and without a sequential generation of heavy
fermions were inferred from those $H\gamma\gamma$ diagrams were the
photons are directly coupled to loop quarks.
In the case of the ${\mathcal O}(x_t)$ corrections, the effects due to the
flipped sign in the top-quark mass counterterm and the interchange of mass
limits in the proper diagrams incidentally compensate each other, so that
Eq.~(61) in Ref.~\cite{Djouadi:1997rj} agrees with
Refs.~\cite{Fugel:2004ug,Djouadi:1994ge}.
In the case of the ${\mathcal O}(x_F)$ corrections, where no mass limits are
interchanged, the fermion mass counterterms cancel within each isodoublet, so
that Eq.~(62) of Ref.~\cite{Djouadi:1997rj} goes unchanged.


\section{\label{sec::numerics}Discussion}

We now explore the phenomenological implications of our results for
$\Gamma(A^0\to\gamma\gamma)$ and $\Gamma(A^0\to gg)$.
For definiteness, we concentrate on the more likely case of three generations.
As explained in Section~\ref{sec::intro}, we consider a scenario with low to
intermediate values of the Higgs-boson masses and $\tan\beta$ and large values
of the supersymmetric-particle masses, so that the dominant electroweak
two-loop corrections are of relative order ${\mathcal O}(x_t)$.
We adopt the following values for our input parameters \cite{Yao:2006px}:
$G_F=1.166\,37\times 10^{-5}$~GeV$^{-2}$, $\alpha_s(M_Z)=0.1176$,
and $m_t = 170.9$~GeV.
As for the unknown 2HDM input parameters, we assume that $M_{A^0}<160$~GeV and
$2<\tan\beta<10$.
For larger values of $\tan\beta$, our approximation of neglecting the
bottom-quark contributions is likely to break down.

\begin{figure}[ht]
\begin{center}
\begin{tabular}{cc}
\includegraphics[width=0.45\textwidth,viewport=19 20 638 435,clip]{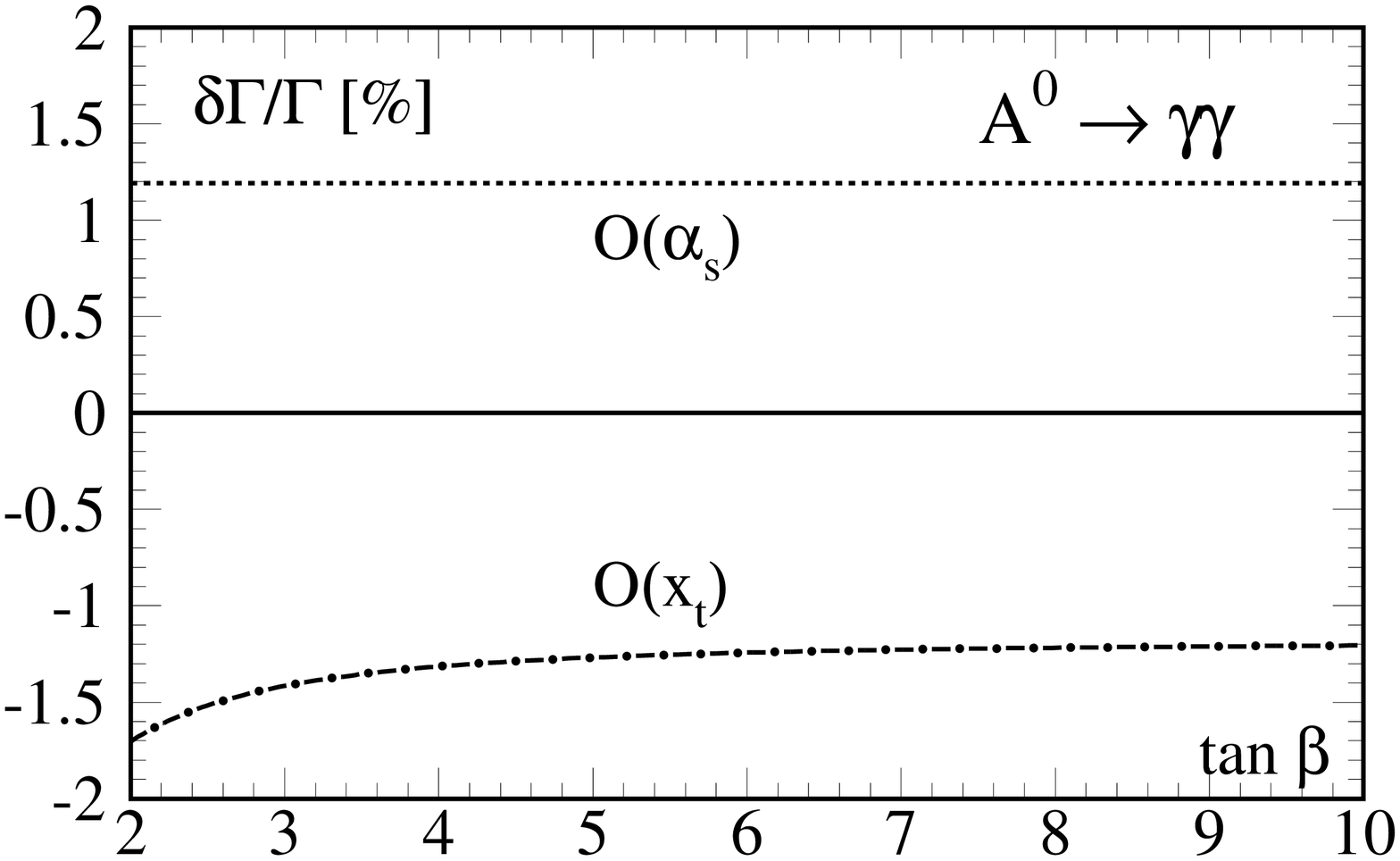}
&
\includegraphics[width=0.45\textwidth,viewport=19 20 638 435,clip]{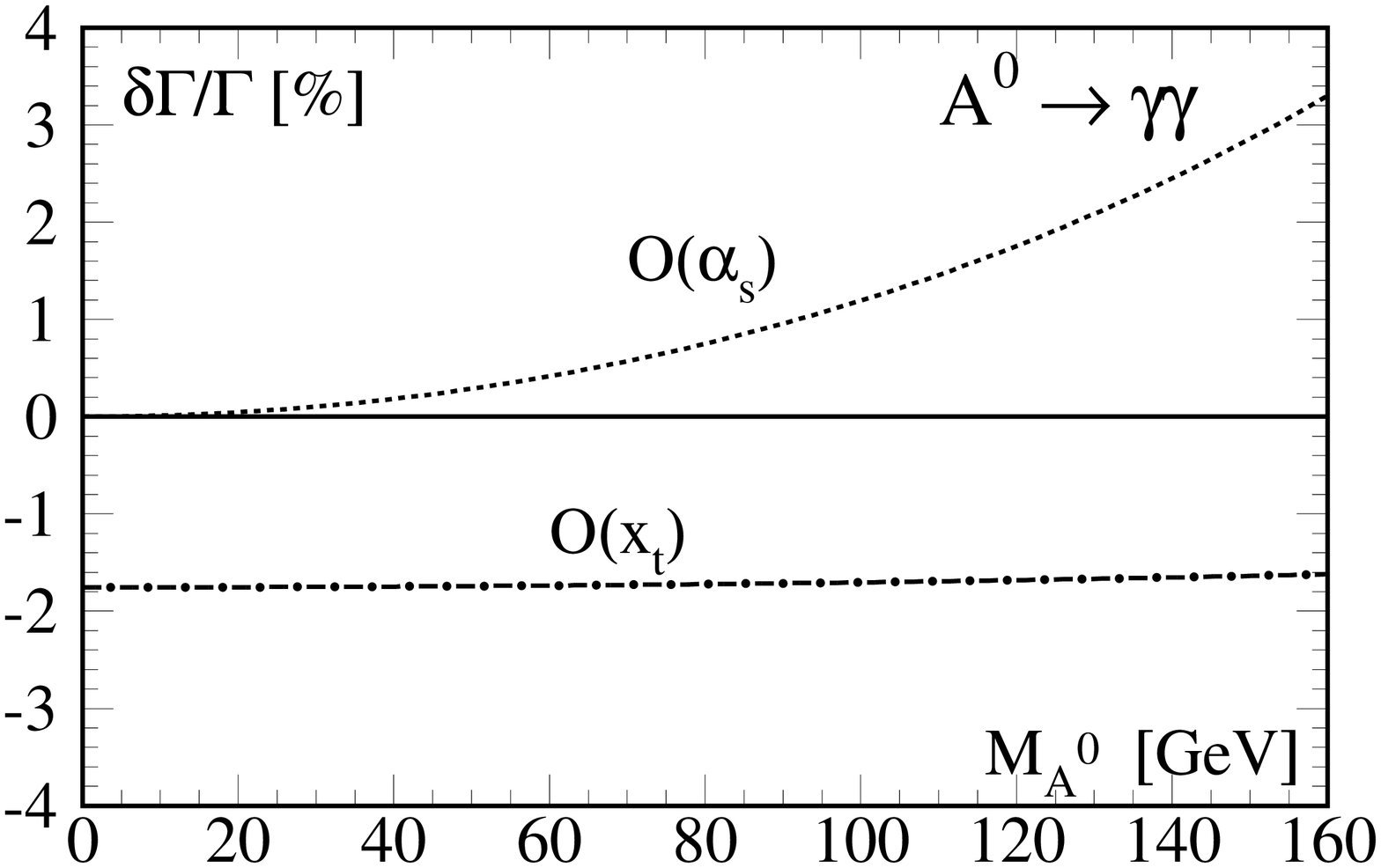}
\end{tabular}
\caption{\label{fig:three}%
${\mathcal O}(x_t)$ and ${\mathcal O}(\alpha_s)$ corrections to
$\Gamma(A^0\to\gamma\gamma)$ (a) for $M_{A^0}=100$~GeV as functions of
$\tan\beta$ and (b) for $\tan\beta=2$ as functions of $M_{A^0}$.}
\end{center}
\end{figure}
We first consider $\Gamma(A^0\to\gamma\gamma)$.
Neglecting the bottom-quark contribution, its ${\mathcal O}(x_t)$ correction
is given by
\begin{eqnarray}
\frac{\delta\Gamma}{\Gamma}&=&
2\frac{\mathcal{A}_t^{x_f}}{\mathcal{A}_t^{\mathrm{LO}}}
\nonumber\\
&=&-x_t\left[4+\frac{7}{\tan^2\beta}+\mathcal{O}(\tau_t)\right],
\label{eq:delgam}
\end{eqnarray}
where, in the second equality, $\mathcal{A}_t^{\mathrm{LO}}$ is approximated
by the leading term in the second line of Eq.~(\ref{eq:res1lA0}) for $f=t$.
We observe that this correction is negative, has its maximum size for small
values of $\tan\beta$, and is independent of $M_{A^0}$, apart from the
$M_{A^0}$ dependence carried by $\mathcal{A}_t^{\mathrm{LO}}$.
Its evaluation according to the first line of Eq.~(\ref{eq:delgam}) is
compared with the ${\mathcal O}(\alpha_s)$ correction in Fig.~\ref{fig:three}.
We observe from Fig.~\ref{fig:three}(a) that the ${\mathcal O}(x_t)$ 
correction amounts to $-1.7\%$ at $\tan\beta=2$ and rapidly reaches its
asymptotic value of $-1.2\%$ as $\tan\beta$ increases, whereas the
${\mathcal O}(\alpha_s)$ correction, evaluated from Eqs.~(\ref{eq:res1lA0})
and (\ref{eq:res2laqcd}), is positive and independent of $\tan\beta$, as long
as the bottom-quark contribution is neglected.
The $M_{A^0}$ dependence of the ${\mathcal O}(x_t)$ correction shown in
Fig.~\ref{fig:three}(b), which is induced by $\mathcal{A}_t^{\mathrm{LO}}$
as mentioned above, is rather feeble, so that we may expect the unknown
${\mathcal O}(\tau_t^n)$ ($n=1,2,3,\ldots$) terms in Eq.~(\ref{eq:delgam}) to
be of moderate size, too.
The smallness and approximately quadratic $M_{A^0}$ dependence of the
${\mathcal O}(\alpha_s)$ correction is due to the absence of the leading
${\mathcal O}(\tau_t^0)$ term in $\mathcal{A}_t^{\alpha_s}$, given by
Eq.~(\ref{eq:res2laqcd}).
We conclude that the ${\mathcal O}(x_t)$ reduction more than compensates the
${\mathcal O}(\alpha_s)$ enhancement for $M_{A^0}\alt120$~GeV.

\begin{figure}[ht]
\begin{center}
\begin{tabular}{cc}
\includegraphics[width=0.45\textwidth,viewport=44 42 647 439,clip]{%
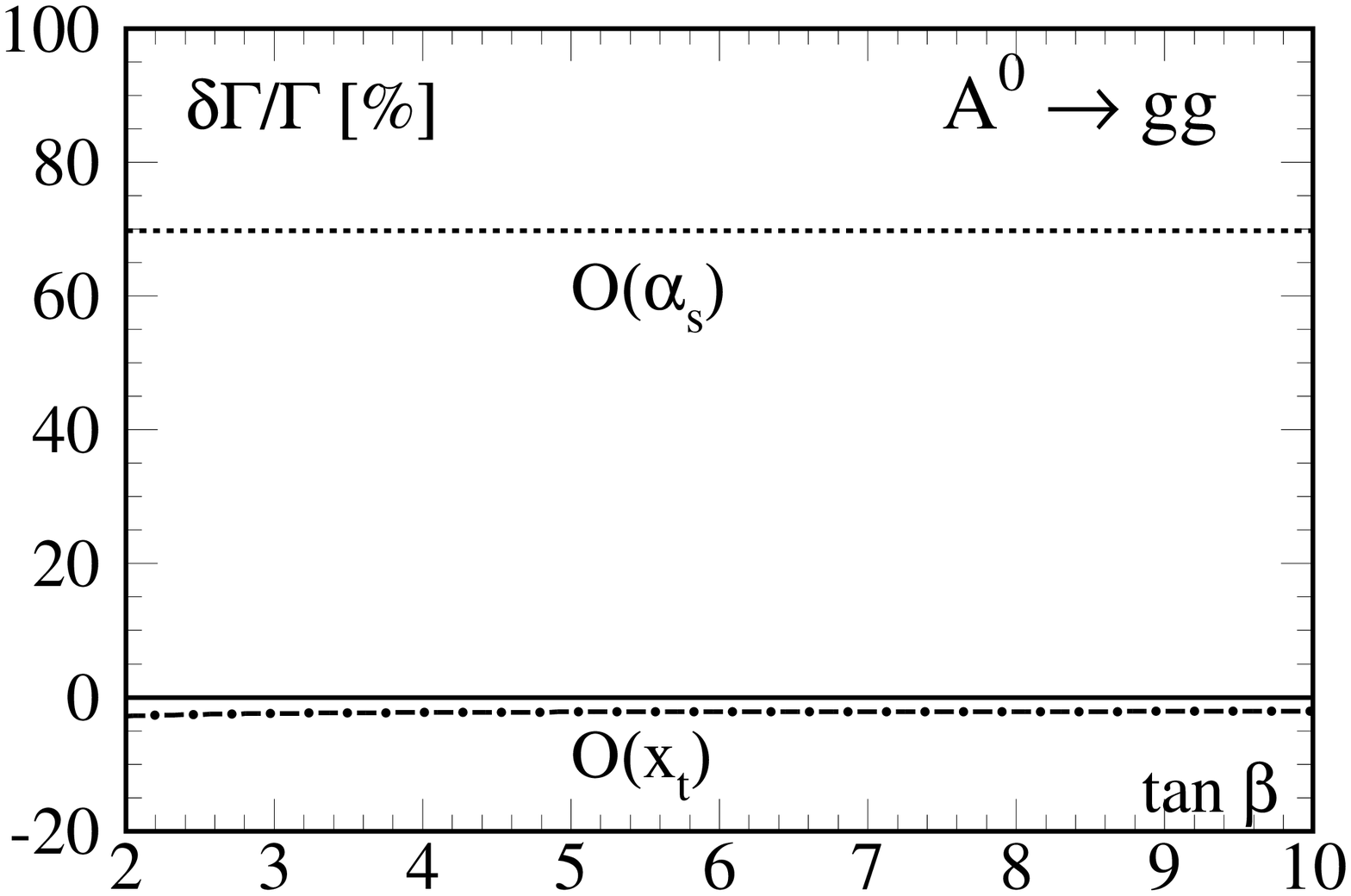}
&
\includegraphics[width=0.45\textwidth,viewport=44 42 647 439,clip]{%
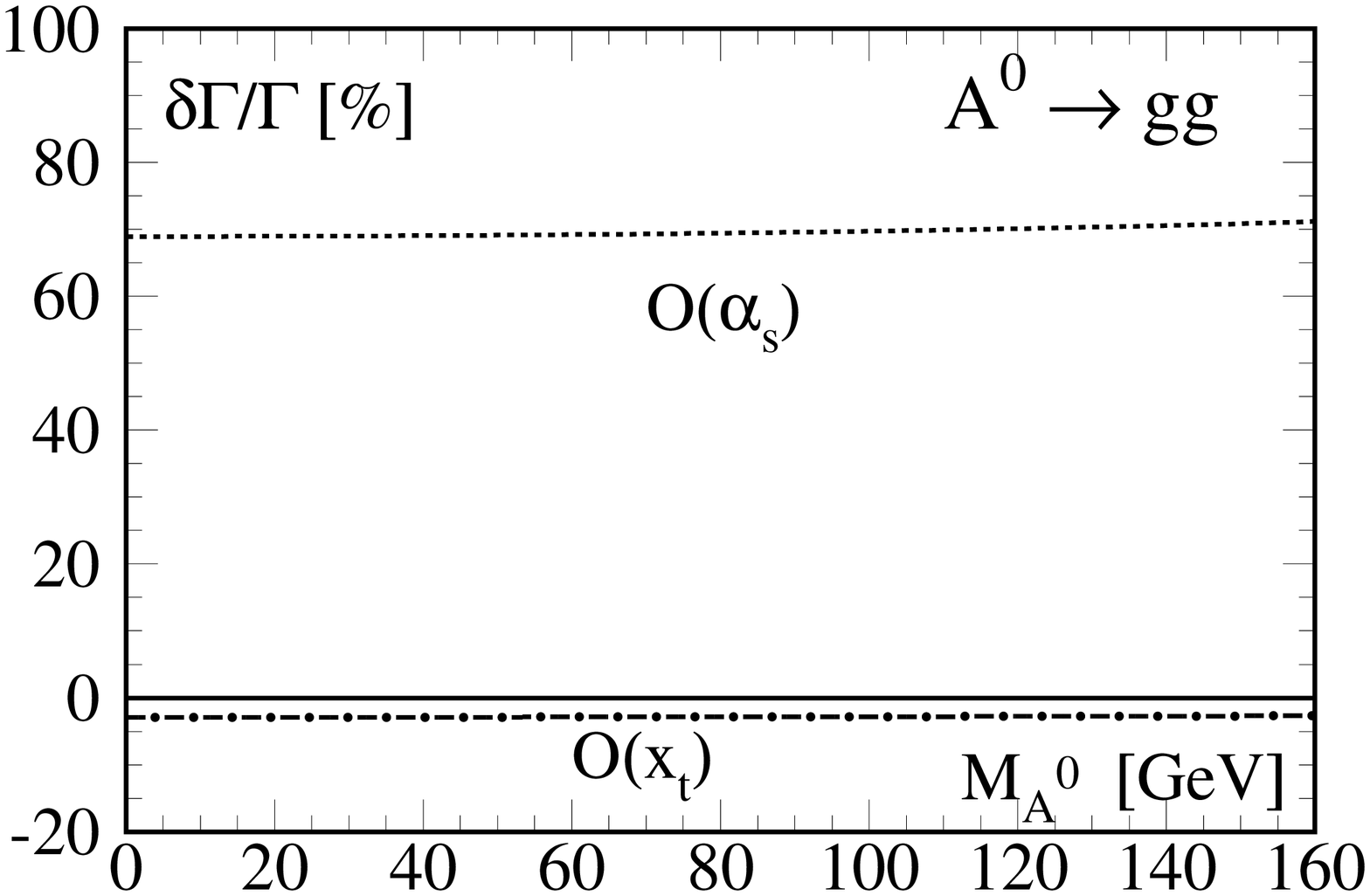}
\end{tabular}
\caption{\label{fig:four}%
${\mathcal O}(x_t)$ and ${\mathcal O}(\alpha_s)$ corrections to
$\Gamma(A^0\to gg)$ (a) for $M_{A^0}=100$~GeV as functions of
$\tan\beta$ and (b) for $\tan\beta=2$ as functions of $M_{A^0}$.}
\end{center}
\end{figure}
We now turn to $\Gamma(A^0\to gg)$.
For $\tau_t\ll1$, its ${\mathcal O}(x_t)$ correction reads
\begin{eqnarray}
\frac{\delta\Gamma}{\Gamma}&=&
2\frac{\mathcal{A}_t^{x_f}}{\mathcal{A}_t^{\mathrm{LO}}}
\nonumber\\
&=&-x_t\left[7+\frac{10}{\tan^2\beta}+\mathcal{O}(\tau_t)\right].
\label{eq:delglu}
\end{eqnarray}
As in the case of $A^0\to\gamma\gamma$, this correction is negative, has its
maximum size for small values of $\tan\beta$, and is independent of $M_{A^0}$,
apart from the $M_{A^0}$ dependence carried by $\mathcal{A}_t^{\mathrm{LO}}$.
In Fig.~\ref{fig:four}, its evaluation according to the first line of
Eq.~(\ref{eq:delglu}) is compared with the full ${\mathcal O}(\alpha_s)$
correction \cite{Spira:1995rr} due to virtual top quarks, which also involves
three-parton final states.
In contrast to the case of $A^0\to\gamma\gamma$, the ${\mathcal O}(\alpha_s)$
correction to $\Gamma(A^0\to gg)$ does have a ${\mathcal O}(\tau_t^0)$ term,
which is about 68\%.
The ${\mathcal O}(\tau_t^0)$ term is also known at ${\mathcal O}(\alpha_s^2)$,
where it is still as large as 23\% \cite{Chetyrkin:1998mw}.
The ${\mathcal O}(x_t)$ correction to $\Gamma(A^0\to gg)$ ranges from $-2.8\%$
at $\tan\beta=2$ to the asymptotic value $-2.1\%$ and partly screens the
sizeable ${\mathcal O}(\alpha_s)$ and ${\mathcal O}(\alpha_s^2)$ corrections.

Let us now briefly comment on the ${\mathcal O}(x_F)$ corrections to
$\Gamma(A^0\to\gamma\gamma)$ and $\Gamma(A^0\to gg)$ due to a sequential
generation of heavy fermions, given in Eqs.~(\ref{eq:UD}), (\ref{eq:NE}),
(\ref{eq:gUD}), and (\ref{eq:gNE}).
These can be sizeable for large values of $m_F$ just because of the prefactor
$x_F$.
For large values of $\tan\beta$, further enhancement comes from the terms of
maximum power in $\tan\beta$, which are cubic for $m_U\approx m_D$,
$m_U\ll m_D$, $m_N\approx m_E$, and $m_N\ll m_E$ in the case of
$A^0\to\gamma\gamma$ and for $m_U\approx m_D$ and $m_U\ll m_D$ in the case of
$A^0\to gg$, and (at most) linear for the other mass hierarchies.
For $\tan\beta>2$, the ${\mathcal O}(x_F)$ corrections reduce the LO results
for $\Gamma(A^0\to\gamma\gamma)$ and $\Gamma(A^0\to gg)$, except for the
corrections due to a $(U,D)$ doublet with $m_U\gg m_D$ and, in the case of
$\Gamma(A^0\to gg)$, also those due to the $(N,E)$ doublet.
Because of the constraint from electroweak precision tests \cite{Yao:2006px}
on the rho parameter \cite{Djouadi:1997rj,Ross:1975fq}, the case of
approximate mass degeneracy within the $(U,D)$ and $(N,E)$ doublets is
favoured, so that a $\tan^3\beta$-enhanced screening is likely to be
encountered.


\section{\label{sec::summary}Conclusions}

In conclusion, we analytically calculated the dominant electroweak two-loop
corrections, of order ${\cal O}(x_t)$, to $\Gamma(A^0\to\gamma\gamma)$,
$\Gamma(A^0\to gg)$, $\sigma(\gamma\gamma\to A^0)$, and $\sigma(gg\to A^0)$
within the 2HDM with low- to intermediate-mass Higgs bosons for small to
moderate value of $\tan\beta$ using asymptotic expansion in
$M_{A^0}^2/(2m_t)^2$.
We also studied how these corrections are modified by the presence of a
sequential generation of heavy fermions, with generic mass $m_F$, and provided
the ${\cal O}(x_F)$ corrections arising then in addition.
We also revisited the ${\cal O}(x_t)$ and ${\cal O}(x_F)$ corrections to
$\Gamma(H\to\gamma\gamma)$ and $\sigma(\gamma\gamma\to H)$ in the
four-generation SM and clarified an inconsistency in
Ref.~\cite{Djouadi:1997rj}.

We recovered the notion that the na\"\i ve treatment of the $\gamma_5$ matrix
being anticommuting in $d$ space-time dimensions leads to ambiguous results,
which depend on the way of executing the Dirac traces.
To consistently overcome the non-trivial $\gamma_5$ problem of dimensional
regularisation, we adopted the HVBM scheme \cite{tHooft:1972fi} and included a
finite renormalisation constant, $Z_5^p$, for the pseudoscalar current to
effectively restore the anticommutativity of the $\gamma_5$ matrix
\cite{Trueman:1979en,Collins:1984xc,Larin:1993tq}.
The ${\cal O}(x_t)$ and ${\cal O}(x_F)$ terms of $Z_5^p$ were found to vanish.
We worked in the electroweak on-shell renormalisation scheme
\cite{Sirlin:1980nh,Hollik:2002mv} endowed with the DCPR definition of
$\tan\beta$ \cite{Chankowski:1992er,Dabelstein:1995js}.

On the phenomenological side, the ${\cal O}(x_t)$ correction to
$\Gamma(A^0\to\gamma\gamma)$ and $\sigma(\gamma\gamma\to A^0)$ is of relative
importance, since it more than compensates the ${\mathcal O}(\alpha_s)$
enhancement for $M_{A^0}\alt120$~GeV.
It leads to a reduction of the LO results, which ranges between $-1.7\%$ and
$-1.2\%$ for $2<\tan\beta<10$ and is independent of $M_{A^0}$.
Such an effect might be measurable for $\sigma(\gamma\gamma\to A^0)$ at the
ILC operated in the $\gamma\gamma$ mode on the $A^0$-boson resonance and
possibly also for $\Gamma(A^0\to\gamma\gamma)$ at the ILC in the regular
$e^+e^-$ mode \cite{Battaglia:2000jb}.

As for $\Gamma(A^0\to gg)$ and $\sigma(gg\to A^0)$, the ${\cal O}(x_t)$
correction screens the sizeable QCD enhancement, by between $-2.8\%$ and
$-2.1\%$ for $2<\tan\beta<10$, and is independent of $M_{A^0}$.
Such a reduction of $\sigma(gg\to A^0)$ should matter at the high luminosities
to be achieved at the LHC.
E.g., given an annual luminosity of 100~fb$^{-1}$ per LHC experiment, a
$pp\to A^0+X$ cross section of about 35~pb (for $M_{A^0}=100$~GeV and
$\tan\beta=2$) \cite{Djouadi:2005gj} amounts to $7\times10^6$ $A^0$ bosons
per year, 2.8\% of which still corresponds to a substantial subsample of
200.000 $A^0$ bosons per year.
Furthermore, the size of this correction is in the ballpark of the theoretical
uncertainty due to the parton distribution functions \cite{Spira:1995rr} and
the scale dependence of the NNLO QCD prediction \cite{Harlander:2002vv}.


\bigskip
\noindent
{\bf Acknowledgements}
\smallskip

We would like to thank M. Gorbahn for helpful discussions,
W. Hollik for a useful communication concerning the renormalisation of
$\tan\beta$,
P. Kant for providing us with the series expansion of the analytic result for
the ${\mathcal O}(\alpha_s)$ correction to $\Gamma(A^0\to\gamma\gamma)$ in
Eq.~(2.20) of Ref.~\cite{Harlander:2005rq},
M. Spira for providing us with the numerical results for the full
${\mathcal O}(\alpha_s)$ correction to $\Gamma(A^0\to gg)$ \cite{Spira:1995rr}
shown in Fig.~\ref{fig:four}, and
M. Steinhauser for providing us with an updated version of \texttt{MATAD}
\cite{Steinhauser:2000ry}.
This work was supported in part by BMBF Grant No.\ 05~HT6GUA and DFG Grant
No.\ GRK~602.


\end{document}